\documentclass[prx,aps,twocolumn,10pt,longbibliography]{revtex4-2}
\usepackage[utf8]{inputenc}
\usepackage{braket}
\usepackage{amsmath}
\DeclareMathOperator{\Tr}{Tr}
\usepackage{amssymb}
\usepackage{mathtools}

\usepackage{graphicx}
\usepackage[shortlabels]{enumitem}
\setlist{nosep}
\usepackage{amsthm}
\usepackage[hidelinks,breaklinks]{hyperref}

\newtheoremstyle{customdefi}{\topsep}{0.3\topsep}{}{}{\bfseries}{.}{.5em}{}
\theoremstyle{plain}

\theoremstyle{customdefi}
\usepackage{algorithm,algorithmicx,algpseudocode}
\usepackage{xcolor}
\usepackage{makecell}
\usepackage{tikz-qtree}
\usepackage{booktabs}
\usepackage{xspace}
\usepackage{subcaption}

\begin{document}
\author{Christoffer Hindlycke}
\email{christoffer.hindlycke@liu.se}
\author{Jakov Krnic}
\email{jakov.krnic@liu.se}
\author{Jan-{\AA}ke Larsson}
\email{jan-ake.larsson@liu.se}
\affiliation{
Department of Electrical Engineering,
Link\"oping University 
581 83 Link\"oping, SWEDEN%
}
\date{\today}
\title{Practical implementation of Toffoli-based qubit rotation}

\begin{abstract}
	The Toffoli gate is an important universal quantum gate, and will alongside the Clifford gates be available in future fault-tolerant quantum computing hardware.
	Many quantum algorithms rely on performing arbitrarily small single-qubit rotations for their function, and these rotations may also be used to construct any unitary from a limited (but universal) gate set. 
	How to carry out such rotations is then of significant
	interest.
	In this work, we evaluate the performance of a recently proposed single-qubit rotation algorithm using the Clifford plus Toffoli gate set by implementation of a one-shot version on both a real and a simulated quantum computer.
  	We test the algorithm under various simulated noise levels using a per-qubit depolarizing error noise model and examine how the probabilities and process fidelities are affected. 
  	We then conduct live runs and find that the results reasonably match the simulated results.
  	We also attempt to model the hardware noise by combining a number of noise models, matching the results to results of the live runs to approximate the hardware noise.
  	Our results suggest that the algorithm will perform well for up to $1\%$ noise, under the noise models we chose.
  	We further posit the use of our algorithm as a benchmark for quantum processing units, given that it has a low complexity that is easy to fine-tune in small steps. We provide details for how to do this.
\end{abstract}
\maketitle

\section{Introduction}
The Toffoli gate is instrumental for quantum computing, being an indispensable part of, for example, Shor's algorithm \cite{Shor1994,Shor1997a}, quantum error correction \cite{Shor1995,Cory1998, Knill2001, Chiaverini2004, Reed2012, Nigg2014}, and shallow quantum circuits \cite{Bravyi2018}. 
Many of these quantum algorithms also require arbitrarily small single-qubit rotations, making approximation of generic gates using a finite gate set essential.
This is especially true in the fault tolerant quantum computing (FTQC) era, where direct fault-tolerant constructions are typically available only for a limited number of gates, so short generation of generic gates is necessary for high-performance computing \cite{Preskill1998}.

The Clifford plus Toffoli gate set will be available in future quantum computer hardware, since the Toffoli gate is indispensable for quantum computing, and a large amount of work has been done toward its practical implementation \citep{Lanyon2009,Monz2009,DiCarlo2009,Fedorov2012,Levine2019}.
Generic unitaries can be built with that gate set \cite{Nielsen2010}; it is therefore natural to ask how this can be done efficiently and explicitly.
We were recently able to give an explicit construction for approximating any $z$-axis rotation to within distance $\epsilon$ using the Clifford plus Toffoli gate set with expected Toffoli count $4\lceil\log\frac1\epsilon\rceil$, see Ref.~\cite{Hindlycke2024a}.
Three such rotations allow the approximation of any single-qubit gate.
Our construction is simple and efficient, requiring only a trigonometric expression of the desired rotation angle, rounded to the desired accuracy.
The resulting algorithm succeeds in applying a rotation $\epsilon$-close to that desired with probability strictly greater than $1/2$, has expected gate depth $4\lceil\log\frac1\epsilon\rceil+6$, and uses $2\lceil\log\frac1\epsilon\rceil$ ancillas, where ancilla denotes an ancillary (helper) qubit.
Should the rotation not be applied, a $Z$ gate is applied instead, and ancilla measurements indicate this.

Beyond its usefulness in an FTQC device, our algorithm also demonstrates a linear scaling in its requisite number of ancilla and Toffoli gates.
This paves the way for it to be used as a remarkably simple means of benchmarking for comparing noisy intermediate scale quantum (NISQ) devices.
The number of qubits in active use and the number of universal gates are typically considered the greatest contributors of noise within such a device \cite{Preskill1998}.
Therefore, if a NISQ device has a native single-qubit rotation available, that rotation would naturally be preferable to the use of the construct studied here.
But the present construction is still useful as a benchmark since its performance is easy to check and the circuit size can be easily adjusted.

The most straightforward way to use our algorithm is to first select a tolerable error rate $\epsilon$ and from that calculate the lowest possible number $n$ of ancillary controls.
The theoretical error rate $\epsilon$ applies in the ideal situation, and an increased ancilla count $n$ would lower~$\epsilon$.
However, with noisy gates a larger gate array size will generate more noise.
The total number of ancillary qubits in active use is $2n - 2$, which also equals the total number of required Toffoli gates.
In other words, increasing $n$ by $1$ increases the total number of qubits and Toffoli gates by $2$ in each case.
To benchmark such an implementation, observe the actual probability of the rotation being applied as compared to the ideal probability, or one could reconstruct the noisy channel via quantum state \cite{Altepeter2004} and process \cite{Chuang1997} tomography and compare it against the ideal unitary, or one could do both of these in tandem.
We provide an explicit example of how to do the latter.

In this work, we first provide a brief description of the single-qubit rotation algorithm from Ref.~\cite{Hindlycke2024a}, and then describe a number of circuit simplifications possible, several of which are new, to our knowledge.
We provide explicit circuits implementing a one-shot version of the algorithm for a number of different rotation angles and numbers of ancillary controls $n$, and then run these circuits within IBM's QISKIT software suite \cite{JavadiAbhari2024}, both using simulated noise in QISKIT's AerSimulator, and using live runs on the QISKIT quantum processing unit (QPU) \texttt{ibm\_torino} \cite{IBMQuantum}.
We then apply quantum state and process tomography to analyze the effect of the quantum channel realized by our (noisy) quantum circuit.
Full experimental details and results are given.
This is followed by a presentation of our results, and some discussion.
Throughout this work, we assume a working knowledge of quantum computation, to which a comprehensive introduction may be found in Ref.~\cite{Nielsen2010}.

\section{The single-qubit rotation algorithm}
\begin{figure}[tb]
	\centering
	a) \includegraphics[scale=1]{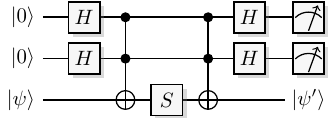}\bigskip
	
	b) \includegraphics[scale=1]{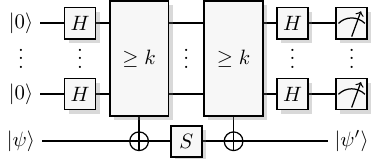}\smallskip
	
	\caption{Circuits for approximate single-qubit rotation. a) Circuit that applies $R_\varphi$ to $\ket{\psi}$ ($\cos \varphi = 3/5$, $\sin\varphi=4/5$) with probability $5/8$; otherwise a $Z$ gate. b) Circuit that applies $R_{\theta^\ast}$ to $\ket{\psi}$, $\epsilon$-close to a desired rotation $R_\theta$ with high probability, where the number of ancillary controls $n$ and the comparison constant $k$ should be chosen as in Eqns.~(\ref{Eq:n})-(\ref{Eq:k}).}
	\label{fig:qsubroutine}
\end{figure}

Let $R_\theta$ denote a single-qubit rotation of angle $\theta$ radians around the $z$ axis of the Bloch sphere.
Any such rotation can be approximated with a circuit that uses a number of ancillary qubits with computational basis initialization and measurements, a number of Toffoli, Pauli-$X$, and Hadamard gates, and finally a single $S=R_{\pi/2}$ rotation~\cite{Hindlycke2024a}.
A simple example (see p.~198 in Ref.~\cite{Nielsen2010}) of a circuit implementing the rotation $R_{\arctan 4/3}$ can be found in Figure~\ref{fig:qsubroutine}a.
The rotation has been performed if the two measurement outcomes both are 0; otherwise the rotation $Z=R_\pi$ has been performed.

That circuit can then be generalized as depicted in Figure~\ref{fig:qsubroutine}b, noting that the Toffoli gate in Figure~\ref{fig:qsubroutine}a is equivalent to a greater than or equal to 3 test on the two control qubits.
Increasing the number of ancillary controls $n$ then allows us to test other $k$ values.
Say we wish to apply a specific rotation of angle $\theta$ to within distance $\epsilon$.
Then by Theorem 1 in Ref.~\cite{Hindlycke2024a}, we should choose the number of ancillary controls 
\begin{equation}
	\label{Eq:n}
	n = 1 + \Bigl\lceil \log \tfrac1\epsilon \Bigr\rceil,
\end{equation} 
and the comparison constant
\begin{equation}
	\label{Eq:k}
	k = 2^{n-1} + \Bigl\lfloor 2^{n-1} \tan\tfrac{\theta}{2} + \tfrac12 \Bigr\rfloor.
\end{equation}
With these values fixed, we construct a circuit carrying out a rotation $R_{\theta^*}$ where
\begin{equation}
	\theta^*=2\arctan\Bigl(\frac k{2^{n-1}}-1\Bigr),
\end{equation}
and we are guaranteed that
\begin{equation}
	||R_\theta- R_{\theta^\ast}|| \leq \lvert \theta - \theta^\ast \rvert \leq \epsilon.
\end{equation}
The theorem also tells us that the probability of successfully applying $R_{\theta^*}$ is greater than $\tfrac12$, heralded by all ancillary controls measuring~$0$.
On failure the $Z$ gate has been applied, which allows easy restoration of the state in the case of failure by simply applying a $Z$ gate on the target qubit output. 

The circuit may then be used in a repeat-until-success algorithm~\cite{Hindlycke2024a}, if so desired,
which requires only two runs on average to perform the target rotation.
Since it is possible to perform conditional loops and if/else clauses for NISQ devices, the algorithm can be easily implemented with the use of current quantum hardware.
In this work, we use the one-shot version of the algorithm; see the chapter on experimental setup (see Section \ref{Sec:ExpSetup}).

\section{Explicit rotation circuit}

\begin{figure}[tb]
	\centering
	\includegraphics[width=\linewidth]{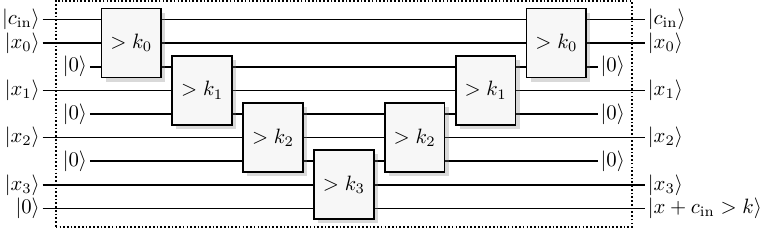}
  a) Comparator circuit from bitwise ripple-carry comparisons\hfill\medskip\\
  \includegraphics{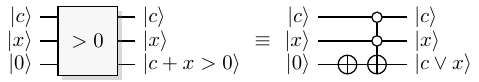}\\
  b) Bit comparison with $k_j=0$; white controls are inverted\hfill\medskip\\
  \includegraphics{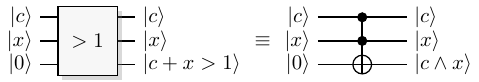}\\
  c) Bit comparison with $k_j=1$\hfill\medskip\\
  \includegraphics{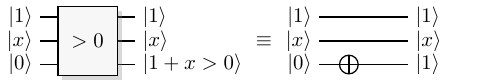}\\
  d) Bit comparison with $k_j=0$ and carry-in $\ket c=\ket1$\hfill\medskip\\
  \includegraphics{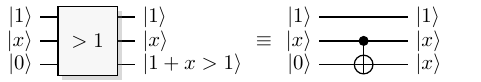}\\
  e) Bit comparison with $k_j=1$ and carry-in $\ket c=\ket1$\hfill\medskip\null
	\caption{Bitwise ripple-carry comparator made of a sequence of Toffoli gates \cite{Hindlycke2024a}. For indices where $k_i$=0, the modified Toffoli gate realizes a quantum OR gate. If carry-in is constant 1, the bit comparisons simplify.}
	\label{fig:ripplecarry}
\end{figure}

The generic circuit performs a comparison with a classical constant $k$ on an ancillary $n$-qubit register.
This is done through a ripple-carry comparator as shown in Figure~\ref{fig:ripplecarry}a, where setting $c_{\text{in}}=1$ gives a $x\ge k$ test.
The individual bit comparisons use one Toffoli gate each, see Figures~\ref{fig:ripplecarry}b and \ref{fig:ripplecarry}c.
We know that $c_{\text{in}}=1$, and then the carry-in can be omitted from the circuit, as can be seen in Figures~\ref{fig:ripplecarry}d and \ref{fig:ripplecarry}e, and there are further simplifications. 

The first simplification consists of reducing the length of $k$ (and thereby $n$) as much as possible.
As long as the least significant bit of $k$ is $0$, the first comparison will check if the input bit sum is greater than or equal to $0$, and so always outputs $1$ for true.
We can then remove that bit from $k$, keep the carry-in of $1$, and reduce $n$ by~$1$.

This happens if $k$ is even, and implies that increasing $n$ is not always beneficial, because different $n$ values may end up giving the same precision.
Subsequently, the first comparison will always be against the bit value $k_0=1$, so the comparison becomes a controlled-NOT gate from $x_0$, the least significant bit of $x$.
We may then simply skip the comparison and use $x_0$ as the first control in the second comparison.
There is one special case: if we reduce all the way to $n=1$ (but not further; this only happens if $k=1$), the approximate rotation $R_\theta^*$ is $R_0=I$, i.e., $\theta^*=0$.
Therefore, the smallest nontrivial $n$ is 2.

The second simplification is that our algorithm applies the greater than or equal to $k$ test, once before and once after applying an $S$ gate to the target, so we may omit the uncomputation part on the first application, and also omit the (forward) computation part on the second application by re-using the ancilla internal to the comparator.
This halves the number of Toffoli gates required.
These simplifications were already used in Ref.~\cite{Hindlycke2024a}.

We now add more simplifications.
The first depends on whether we have native access to an OR gate (Toffoli gate with inverted controls and inverted target) or do not have native access.
If we have access only to a standard Toffoli gate, an OR gate needs to be created with the use of $X$ gates before and after the controls, and before the Toffoli target.
We push the $X$ gates on the left side toward the sources, remembering that an $X$ gate on the target of a Toffoli gate commutes with the Toffoli gate.
Likewise we push the $X$ gates on the right side toward the measurements.
The $X$ gates between the computation and uncomputation part cancel since $X^2=I$.
For the qubits in the comparison register $x$, the $X$ gates have no effect at either the source or the measurement since $X\ket\pm=\pm\ket\pm$, so can be removed. 

For the ancilla internal to the comparator, we can simply absorb the $X$ gate into the state preparation, and there is no need to return the internal ancillas to the state $\ket{0}$, they must simply be returned to a known state.
Note that if two consecutive bits of $k$ are both 1, we will absorb two $X$ gates into the preparation of the internal ancilla that connects the corresponding single-bit comparisons.
This further reduces the circuit complexity.
Similarly, rather than applying an initial set of Hadamard gates to the external ancilla, we can initialize them in the state $\ket{+}$.
Some architectures may even have native support for measuring in different bases. 
If so, we may omit the final $n$ Hadamard gates as well, and opt to measure in the Pauli $X$ basis.

Finally, if the most significant bit of $k$ is 0, the last bit comparison will contain $X$ gates applied to the target qubit, in which case these gates can be absorbed into the $S$ gate to create an $S^\dagger$ gate in its place.

\begin{figure}[tbp]
	\includegraphics[width=\linewidth]{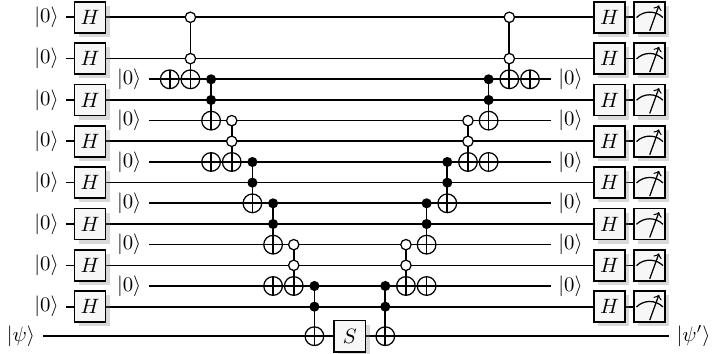}
	\bigskip
	
	\includegraphics[width=\linewidth]{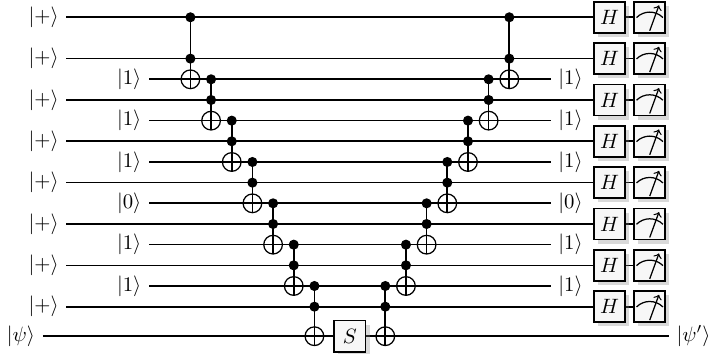}
	
	\caption{Gate array that approximates the $T$ gate to within $\epsilon < 2.6\times10^{-4}$, before and after simplifications, when performing manual basis switching prior to measurement; here $k = 181 = 10110101_2$.}
	\label{fig:ReductionExample}
\end{figure}

We exemplify the above in Figure~\ref{fig:ReductionExample} by comparing our algorithm approximating the $T$ gate for $n = 8$, before and after the additional simplifications described above.
Here we manually switch basis before measuring.
In this case, assuming we consider a Toffoli gate with inverted controls as using a total of four $X$ gates, we were able to reduce the $X$ gate count by $30$, and the Hadamard gate count by $8$.
In general, for a reduced $k$ and $n$, the $X$ gate count reduces by 
$10 \sum_{i} (k_i \oplus 1)$, and the Hadamard count reduces by $n$ ($2n$ if we can measure directly in the Pauli $X$ basis).
This also reduces the circuit depth considerably.

\section{Experimental setup for a tomographic experiment}
\label{Sec:ExpSetup}
The intent behind our experiment is two-fold.
Our primary concern is verifying theoretical predictions as to the behavior of our algorithm by comparing them with experimental outcomes.
Also, we should like some estimation of our algorithm's performance under various noise levels.

To achieve this, we reconstruct the actual channel implemented by the algorithm using quantum tomography.
There are two stages to this.
In the first stage, quantum state tomography is applied to estimate the outcome when the channel our algorithm implements acts on a number of basis states \cite{Altepeter2004}.
In the second stage, these estimates are used to reconstruct the channel in question via quantum process tomography \cite{Chuang1997}.
The reconstructed channel may then be compared with a target unitary.

Our means of this comparison is by the Process Fidelity (PF) \cite{Nielsen2002} which is expressed as
\begin{equation}
  F_{\text{pro}}(\mathcal{E},\mathcal{F})=
	F(\rho_\mathcal{E}, \rho_\mathcal{F}) = \biggl(\Tr\sqrt{\sqrt{\rho_{\mathcal{E}}}\rho_{\mathcal{F}}\sqrt{\rho_{\mathcal{E}}}}\biggr)^2,
\end{equation}
i.e., the state fidelity between the normalized Choi matrix $\rho_\mathcal{E}$ of the quantum channel, calculated with use of quantum process tomography as described in Ref.~\cite{Chuang1997}, and the normalized Choi matrix of the desired unitary $\rho_\mathcal{F}$, calculated directly from the unitary.
We should also mention that Average Gate Fidelity (AGF) \cite{Nielsen2002} is another common way of measuring how ``close'' two unitaries are to each other, and we note that AGF is an affine transformation of PF, and so we use only PF in this paper.

The quantum process tomography in Ref.~\cite{Chuang1997} requires (approximations of) the output state from the unknown channel acting on the basis states $\ket{0}$, $\ket{1}$, $\ket{+}$, and $\ket{+i}$.
Approximating this in turn requires preparing each of these states, applying the channel, and then measuring each state in the three Pauli bases $X$, $Y$, and $Z$.
Hence, for a fixed number of external ancilla $n$, we need a total of $12$ circuits.

For the simulated runs, in addition to perfect (error-free) runs, we also introduce simulated errors occurring when applying a gate and when measuring.
We run simulations using a number of error types available in QISKIT - namely, depolarizing errors, phase damping errors, and their combination - to compare their respective effects on the algorithm.
We apply both single-qubit and Toffoli errors since single-qubit and Toffoli gates are used in the algorithm, and adjust the error levels to estimate the noise in the machine.
Note that error-free runs will yield a unitary to within statistical precision, while all other runs yield a more generic channel.

For the live runs, in addition to running the circuits directly, error mitigation is also utilized in the form of (separately applied) dynamical decoupling and Pauli twirling.
Dynamical decoupling \cite{Viola1999} means adding Pauli gates, whose effect sum to the identity, to prevent qubits from sitting idle for too long.
Since we use a ripple-carry comparator, it is indeed the case that most of our ancilla qubits are idle throughout most of the execution time.
Owing to the technical implementation of QISKIT QPUs, this may then help mitigate errors.
Pauli twirling \cite{Bennett1996} entails adding Pauli gates before and after the gates of a circuit in such a way that the effect of each gate remains the same.
The intent here is to transform the (unknown) noise of the channel into Pauli noise, which tends to accumulate more slowly \cite{Shor1997b}.
Pauli twirling can be combined with error mitigation techniques specifically aimed at suppressing Pauli noise \cite{Shor1997b}, although we have not done so here.
Live runs were performed with the QISKIT QPU \texttt{ibm\_torino}, while simulated runs were performed with QISKIT's AerSimulator.

While there exists a repeat-until-success version of the single-qubit rotation algorithm in Ref.~\cite{Hindlycke2024a}, we elected to use a one-shot version, meaning that the number of repetitions is set to $1$.
We ran the algorithm for $2 \leq n \leq 8$ external ancilla, because $n=2$ corresponds to the smallest nontrivial circuit.
The $n$ reduction at even $k$ will cause some of the data points to coincide; we indicate this in the figure captions.

All runs were used to build a point estimate of the probability of the all-zero outcome of the ancillary controls, the cases that indicate successful application of $R_{\theta^*}$.
On the basis of whether $R_{\theta^*}$ or $Z^*$ was applied, measurements of the target qubit were then used to approximate the two unitaries using quantum state and process tomography.
The output is two quantum channels that also contain noise.
With these estimates in hand for a given $n$, we then calculated the PF \cite{Nielsen2002} compared with that for the $R_\theta$ gate (when $R_{\theta^*}$ was applied) and the $Z$ gate (when $Z^*$ was applied).

\section{Generic behavior through simulation}

In this section we present the results and analysis of simulated and live runs. 
Section \ref{sec:arbitrary_angles} covers simulation of arbitrary rotation angles for differing amounts of depolarizing error. 
We use the results to draw conclusions on how the algorithm behaves for different error rates. 
In Section \ref{sec:simulation_Trotation} we focus on results when applying a $T$ rotation and analyze how the probabilities change for different depolarizing errors and $n$ values. 
We also discuss how the probabilities for a specific rotation angle $\theta$ are connected to the probabilities for arbitrary rotations covered in Section \ref{sec:arbitrary_angles}. 
Finally, in Section \ref{sec:simulation_PF} we study how the PF of the approximation is influenced by circuit size.

\subsection{Circuit noise dependence on rotation angle}
\label{sec:arbitrary_angles}

\begin{figure}[tbp]
	\includegraphics[width=\linewidth,trim=125 25 125 114,clip]{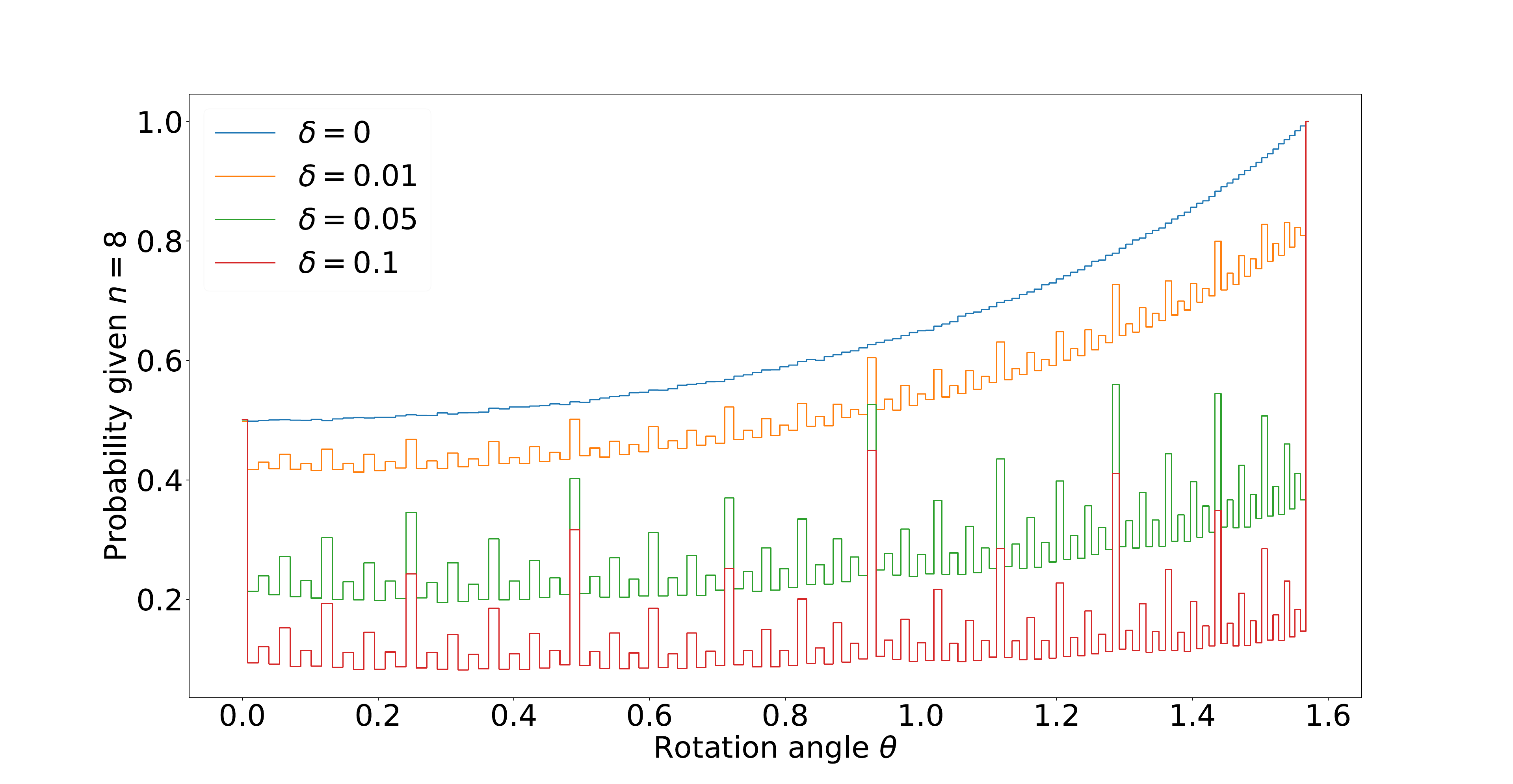}
	\caption{Rotation probabilities for different angles $\theta$, given $n$ = 8, for which $|\theta - \theta^*|$ = 0. The bars arise as the algorithm periodically reduces $n=8$ to a lower value, reducing the circuit complexity and thereby the errors. For each error rate the bar at $\theta=2\arctan\frac12\approx 0.93$ rises highest above the baseline and corresponds to $n=2$, while the two next-highest bars are at $\theta=2\arctan\frac14\approx0.49$ and $\theta=2\arctan\frac34\approx1.29$ and correspond to $n=3$, and so on.}
	\label{fig:ExactAnglesProb}
\end{figure}

Since the algorithm depends in a complicated manner on the rotation angle, we first illustrate the behavior in some simulated runs using a relatively simple noise model.
In this subsection we study the probability of successfully applying an $R_\theta$ rotation for $\theta$ in the interval $[-\frac{\pi}{2},\frac{\pi}{2}]$ and a maximum $n$ of $8$; see Figure~\ref{fig:ExactAnglesProb}.
Each data point corresponds to $10^4$ circuit iterations.
The noise used is simulated depolarizing error with error probabilities $\delta$ of 0, 0.01, 0.05, and 0.1.
We plot only positive angles since the error rate would be symmetrical around $\theta=0$.
The difference between positive and negative angles is that the most significant bit of $k$ will be set to 0 or 1, respectively, corresponding to applying either an $S$ gate or an $S^\dagger$ gate to the target qubit as described above.
The algorithm behaves as expected for $\theta=\pm\frac{\pi}{2}$, as it reduces the circuit to $n=0$, corresponding to a single $S$ or $S^\dagger$ gate. 
When $\theta=0$, the array becomes completely trivial, corresponding to the identity.

The curves that include noise resemble bar charts, where certain angles have a much higher probability of being applied than their neighbor.
The reason for the bars is the rounding procedure of the algorithm, which chooses the closest integer $k$ to the value $2^{n-1}+2^{n-1}\tan(\theta/2)$, or in other words constant $\theta^*$ for all $\theta$ in an interval.
The algorithm also reduces $n$ if the lowest significant bits of $k$ equal $0$, both simplifying the circuit and reducing noise, giving the bar chart resemblance, where the different bars correspond to different $k$ values that reduce to different $n$ values.
Higher bars correspond to smaller $n$ values, and $k$ values that generate the same $n$ value have similar noise dependence of the probability.
A clear example of this is the tallest bar, which corresponds to the angle $\theta = \arctan(4/3)=2\arctan(1/2)$, for which the circuit reduces to $k=3$ and $n=2$, corresponding to that in Figure~\ref{fig:qsubroutine}a \cite{Nielsen2010}.

\subsection{Circuit noise dependence on circuit size}
\label{sec:simulation_Trotation}
That the noise also depends on the circuit size is natural and already visible in Figure~\ref{fig:ExactAnglesProb}.
For a given rotation angle, the noise will vary with the maximum value of $n$ used, at least if $k$ does not contain zeros in its least significant bits.
We illustrate this here using the probability of successfully applying the $T=R_{\pi/4}$ gate in some simulated runs, similarly to what was done above.
The algorithm probability of applying a $T^*$ rotation at different $n$ values and different $\delta$ values is shown in Figure ~\ref{fig:PlotSimProb}.

For an $R_{\pi/4}$ rotation, the value $\tan(\theta/2)$ is irrational so there will be a rounding error for all $n$; however, the rounding procedure gives a 0 in the least significant bit of $k$ for $n=3$, $n=9$, $n=10$, $n=11$, and $n=12$, for example, so will reduce the circuit size in those cases.
We therefore plot data only to $n=8$.
Although somewhat difficult to tell from the plot, $\delta = 0$ gives outcomes in excellent accordance with theoretical predictions.
As expected from the increasing gate depth and number of measurements necessary as $n$ increases, increasing $\delta$ leads to decreasing probability of success. 
Even so, at $\delta = 0.01$, we still have probability (close to) $1/2$ of applying a $T^*$ rotation.

Comparing the theoretical predictions with the simulations at the different $\delta$ values, we find that the latter are proportional to the ideal prediction, but with an additional exponential dependence on the number of ancillas~$n$.
For $n\ge4$, the simulated curves at nonzero $\delta$ are all close to
\begin{equation}
  P_\delta(0^n)=P_0(0^n)(1-c_\delta)^{n-1};
  \label{eq:exponential}
\end{equation}
see the dotted lines in Figure~\ref{fig:PlotSimProb}. 
The case $n=2$ (and $n=3$) do not follow this pattern.
One possible reason for this is that there are no ancillas internal to the comparator with $k$, which likely gives a different error propagation.
Equation (\ref{eq:exponential}) suggests that the simulations behave as if all the errors change only controlling ancilla measurement outcomes from 0 into 1, independently with equal probability $c_\delta$ for each controlling ancilla.
The values used in Figure~\ref{fig:PlotSimProb} are $c_{0.01}=0.02925$, $c_{0.05}=0.13280$ and $c_{0.1}=0.23788$, all of which are close to $3\delta$ or actually closer to $3\delta(1-\delta)^2+\delta^3$ (with values 0.02940, 0.13550, and 0.24400), the probability of an odd number of single qubit errors in a Toffoli gate.
The exponent corresponds to the number of Toffoli gates, so this would indicate that Toffoli errors dominate the noise, which corresponds to our expectations given the circuit layout.

It is worth noting a connection between Figures \ref{fig:ExactAnglesProb} and~\ref{fig:PlotSimProb}.
For each value of $n$, the rounding procedure will give a $\theta^*$ that corresponds to one of the bars in Figure \ref{fig:ExactAnglesProb}, depending on the value of $n$ used.
For example, application of $\theta = \frac{\pi}{4}$ with $n=2$ gives $\theta^*\approx 0.93$, corresponding to the high bar at that point in Figure \ref{fig:PlotSimProb}.
Use of $n=3$ gives the same $\theta^*$, while use of $n=4$ corresponds to $\theta^* \approx 0.72$.
Since we move to the left in Figure \ref{fig:ExactAnglesProb} when changing from $n=2$ to $n=4$, the probability is decreased, as seen in Figure \ref{fig:PlotSimProb}.
Changing to $n=5$, we obtain $\theta^*\approx0.82$, so we move to the right, thereby increasing the probability (this trend can be best seen by looking at the graph for $\delta = 0$ in Figure \ref{fig:PlotSimProb}). This can be useful as we can choose to pick a nearby peak (corresponding to a different angle) instead of the given rotation angle, resulting in higher probability of rotation at the expense of a slight deviation in the angle being applied.

\subsection{Process Fidelity (PF) dependence on circuit size}
\label{sec:simulation_PF}
\begin{figure}[tbp]
	\subcaptionbox{Probability of applying $T^*$, simulated runs.\label{fig:PlotSimProb}
  \medskip}{\includegraphics[width=\linewidth,trim=125 25 125 114,clip]{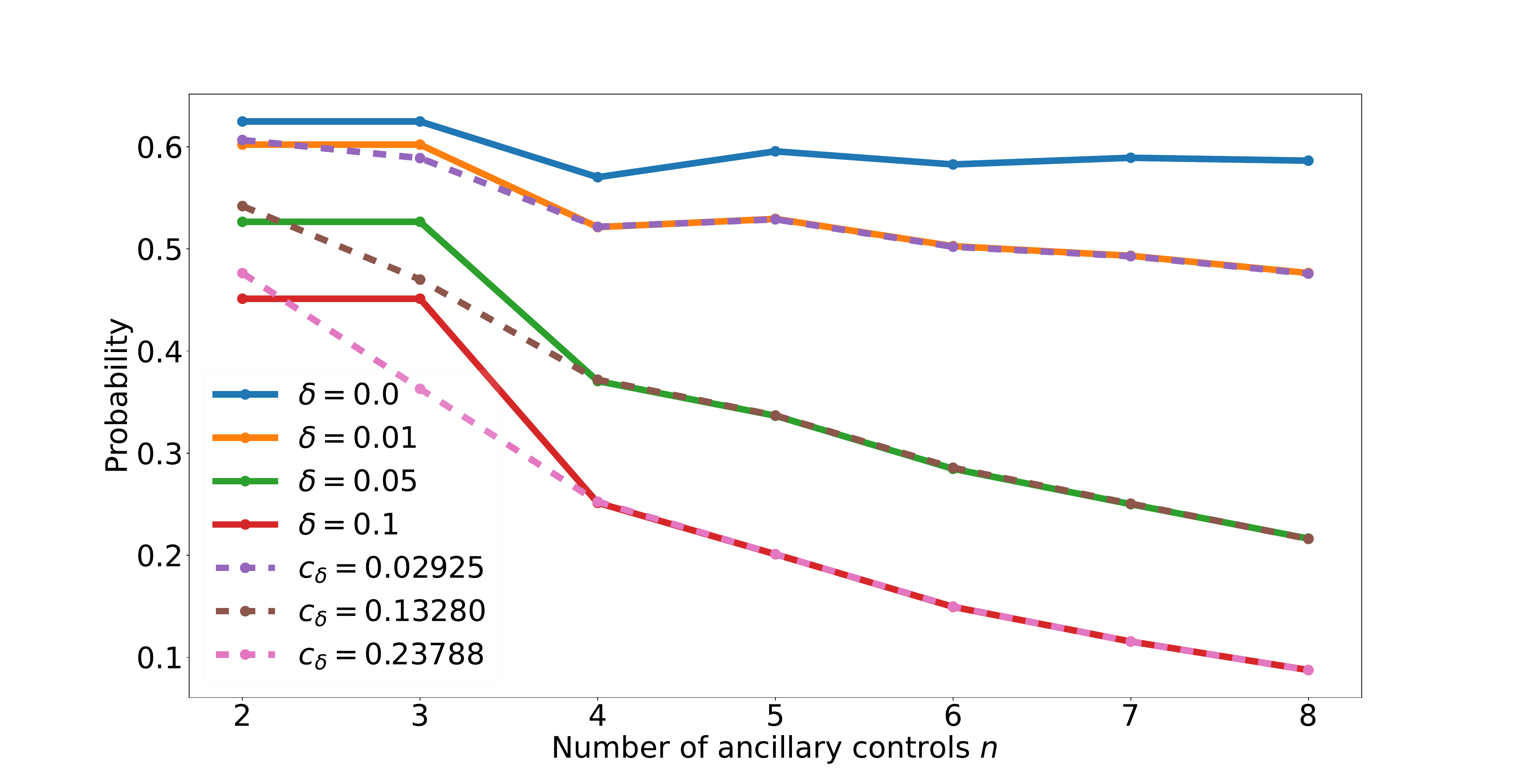}}
	\subcaptionbox{PF calculated for $T^*$, simulated runs.\label{fig:PlotSimPFT}\medskip}{\includegraphics[width=1\linewidth,trim=100 25 125 110,clip]{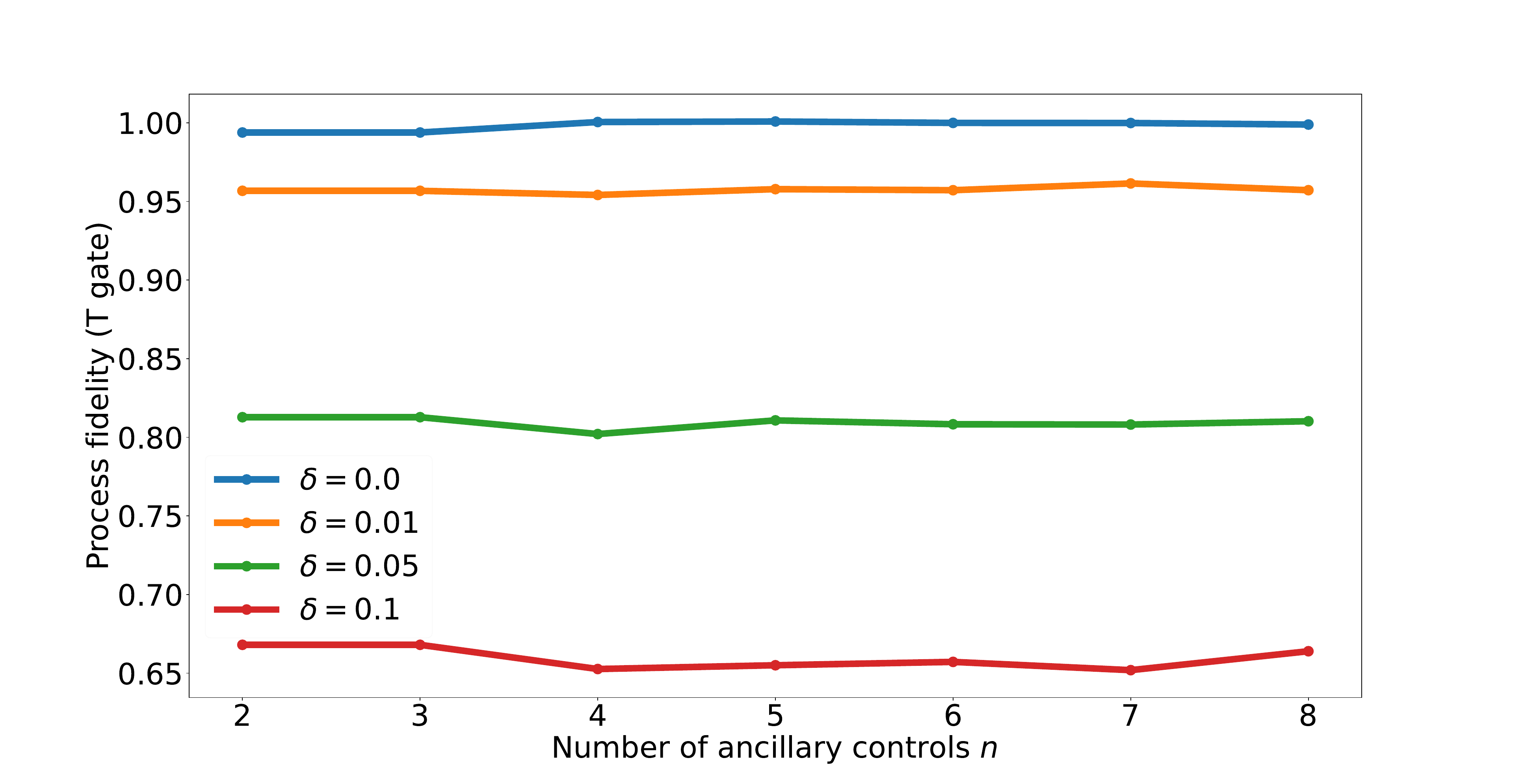}}
	\subcaptionbox{PF calculated for $Z^*$ ($\theta = \pi/4$), simulated runs.\label{fig:PlotSimPFZ}}{\includegraphics[width=1\linewidth,trim=125 25 125 110,clip]{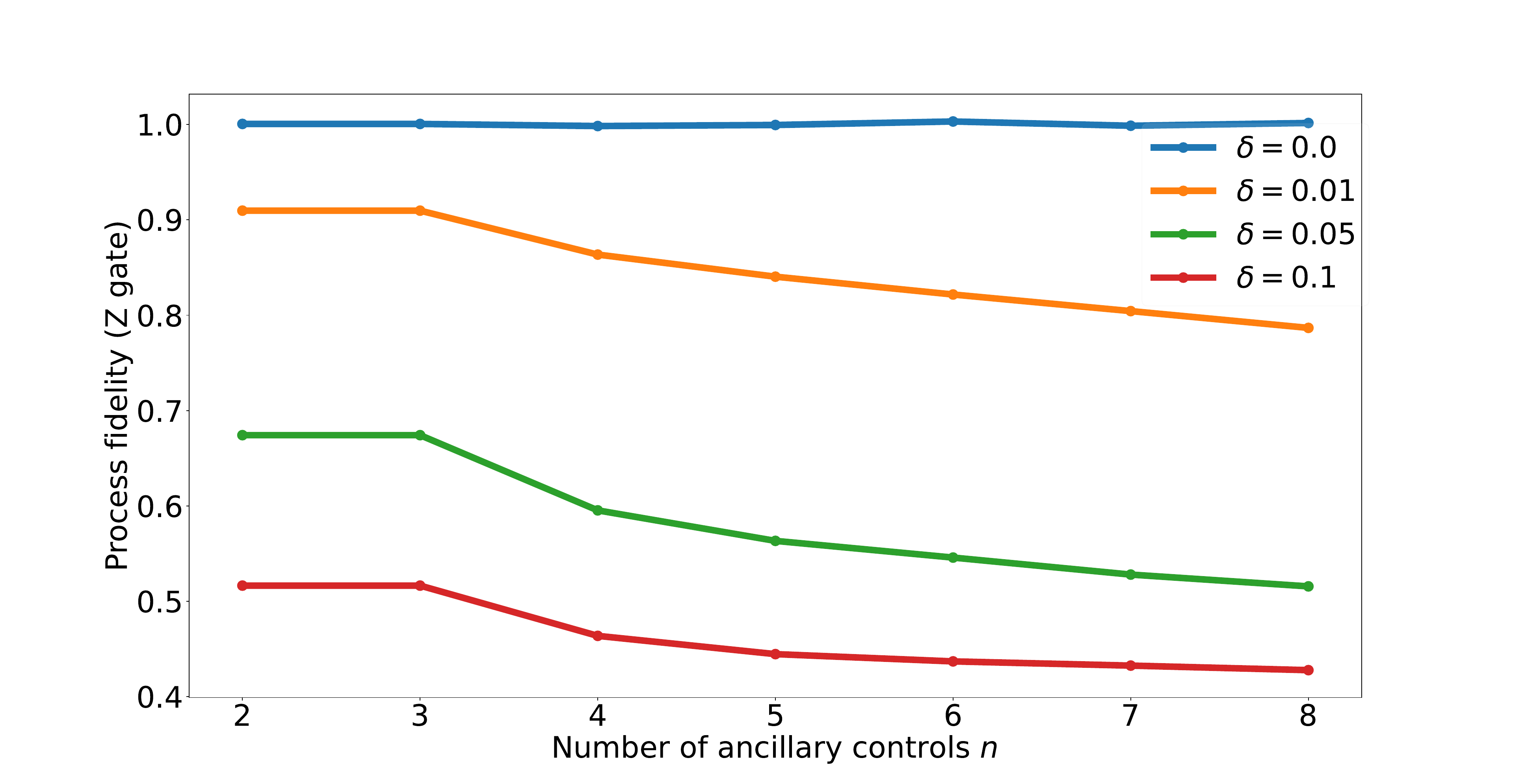}}
	\caption{Performance for the $T^*$ circuit, simulated runs. Since the same circuit is used for $n=2$ and $n=3$, the same data were used.}
  \label{fig:PlotSim}
\end{figure}

When the controlling ancillas are all 0, the circuit should have performed the $T^*$ rotation. 
We use the PF to assess the quality of the rotation, to compare the actual noisy $T^*$ rotation with the desired $T$ gate.
The simulated PF is shown in Figure \ref{fig:PlotSimPFT}.
We point out that the angle error $|\theta-\theta^*|$ is $0.14190$ at $n=2$ (and $n=3$) which may seem large in radians, but corresponds to a PF of $\frac12+\frac12\cos(\theta-\theta^*)=0.99497$.
The increase for $n = 4$ is just visible in Figure \ref{fig:PlotSimPFT}. 
The angle error is $0.06786$, which corresponds to a PF of $0.99775$, which is close enough to 1 to push the simulation output above 1, because of numerical instability in the calculations (this is a well-known problem in quantum tomography); see Table \ref{tbl:prob} in Appendix \ref{app:Tables}.

It is worth noting that the PF for a $T^*$ rotation is hardly affected when $n$ is increased.
Interestingly, the PF remains almost constant as we increase $n$, suggesting that larger values of $\delta$ only influence the ``starting point'' as it were. 
The reason is likely the already noted behavior, that the errors in the ancillary controls will give nonzero ancillary outcomes, so a failure rather than an erroneous rotation that would be counted as a successful rotation.
The errors directly influencing the target will remain, and constitute the reason for the lowered starting point.
When the PF is used as figure of merit, there seems to be little to gain from increasing $n$, and there is a benefit (if any) only for very low noise.
The reason is that already at $n=2$ the PF in the noiseless case is $0.994$; see Table~\ref{tbl:sim} in Appendix~\ref{app:Tables}.

The PF plot for the failed runs, when one or more controlling ancilla outcomes are nonzero, shows a slightly different behavior.
In this case we compare the noisy $Z^*$ rotation with the ideal $Z$ gate, as depicted in Figure \ref{fig:PlotSimPFZ}.
Again, $\delta = 0$ gives simulation outcomes that align with theoretical predictions, but here for nonzero $\delta$ the PF decreases slowly with increased $n$; the rate here is also exponential at about $1\%$-$1.5\%$ decrease per step in $n$.
The reason for the decreasing PF for $Z^*$ rotation is likely that errors in the controls, causing additional nonzero controlling ancilla measurement outcomes, contribute in this case, unlike for~$T^*$ rotation.
Thus, we see errors both from the target and the controls in this case, meaning that an increased number of controls increases the errors.

Given the propensity for gradually decreasing PF between $Z^*$ rotation and the $Z$ gate, it might in the case of systems with more substantial error rates be advisable to stick to the present one-shot version of the algorithm, rather than the repeat-until-success version also found in Ref.~\cite{Hindlycke2024a}, instead compensating by performing a post-selection procedure to retrieve valid measurements.

\section{Live runs and noise benchmarking} \label{sec:live_runs}
Subsection \ref{subsec:live_runs} contains results from live runs performed on \texttt{ibm\_torino}, where we apply a $T^*$ rotation and a $\sqrt{T}^*$ rotation.
The rotation circuits applying the $T^*$ and $\sqrt T^*$ rotations are found in Appendixes \ref{app:Circuits} and \ref{app:CircuitsPi8}.
In Section \ref{sec:sim_live_comparison} we attempt to model the hardware noise by combining three separate error models. Extensive simulations are performed where the error rates for the respective models are varied and fitted to the live run results.

\subsection{Live runs}
\label{subsec:live_runs}
\begin{figure}[tbp]
	\subcaptionbox{Probability of applying $T^*$, live runs.\label{fig:PlotTorinoProb}\medskip}{\includegraphics[width=1\linewidth,trim=110 25 125 114,clip]{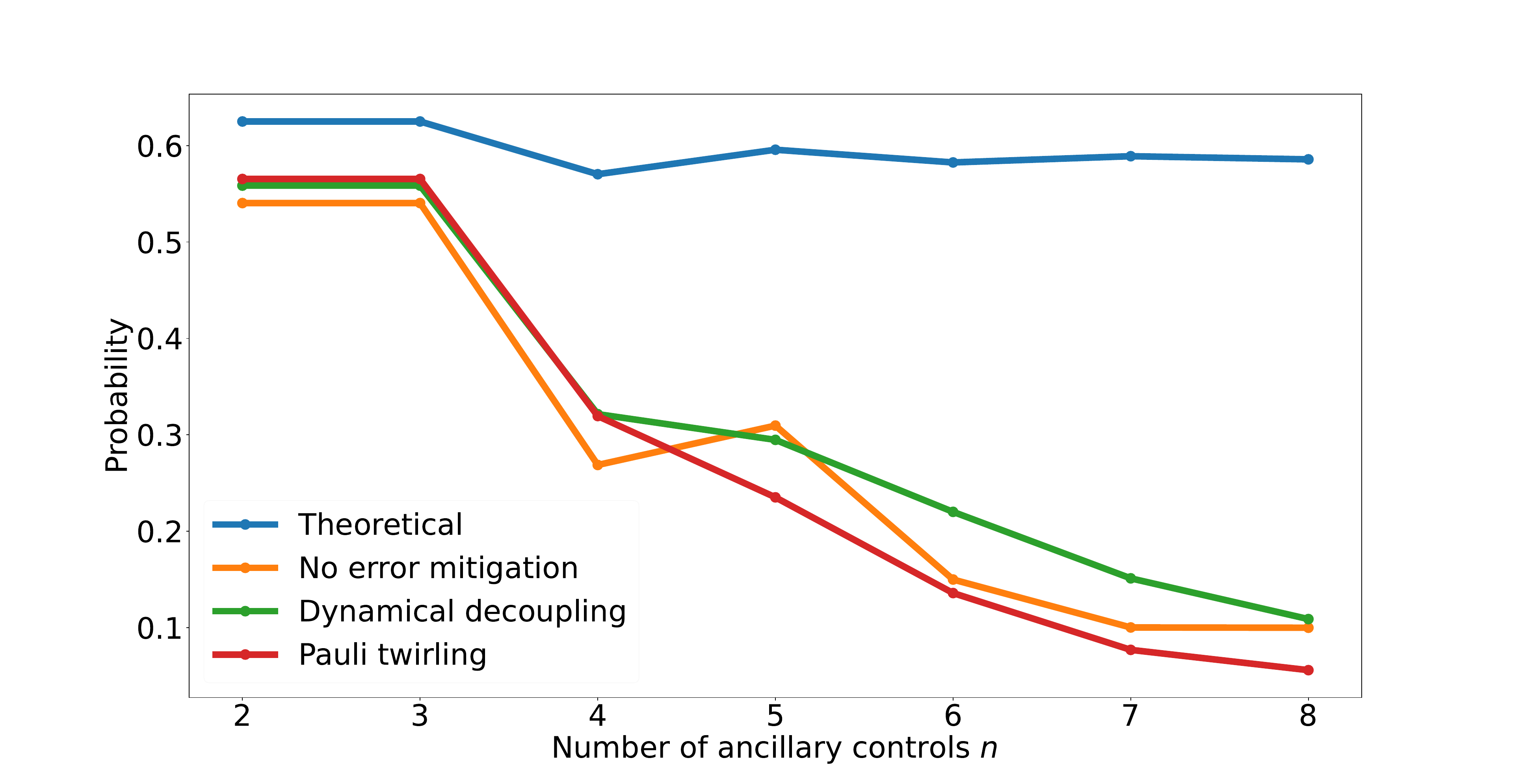}}
  \subcaptionbox{PF for $T^*$, live runs. \label{fig:PlotTorinoPFT}\medskip}{	\includegraphics[width=1\linewidth,trim=100 25 125 114,clip]{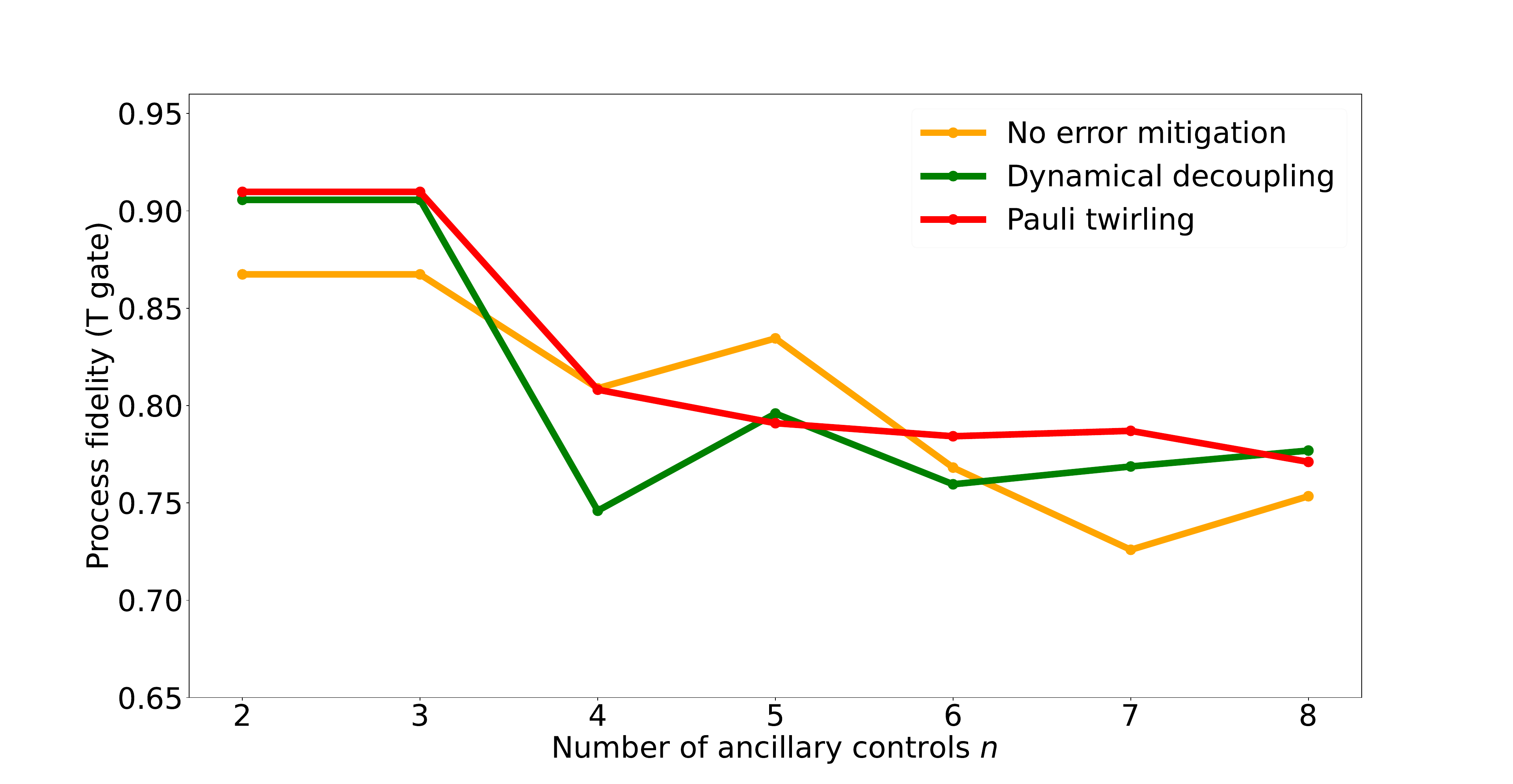}}
  \subcaptionbox{PF for $Z^*$ ($\theta = \pi/4$), live runs.\label{fig:PlotTorinoPFZ}
  }{\includegraphics[width=1\linewidth,trim=110 25 125 114,clip]{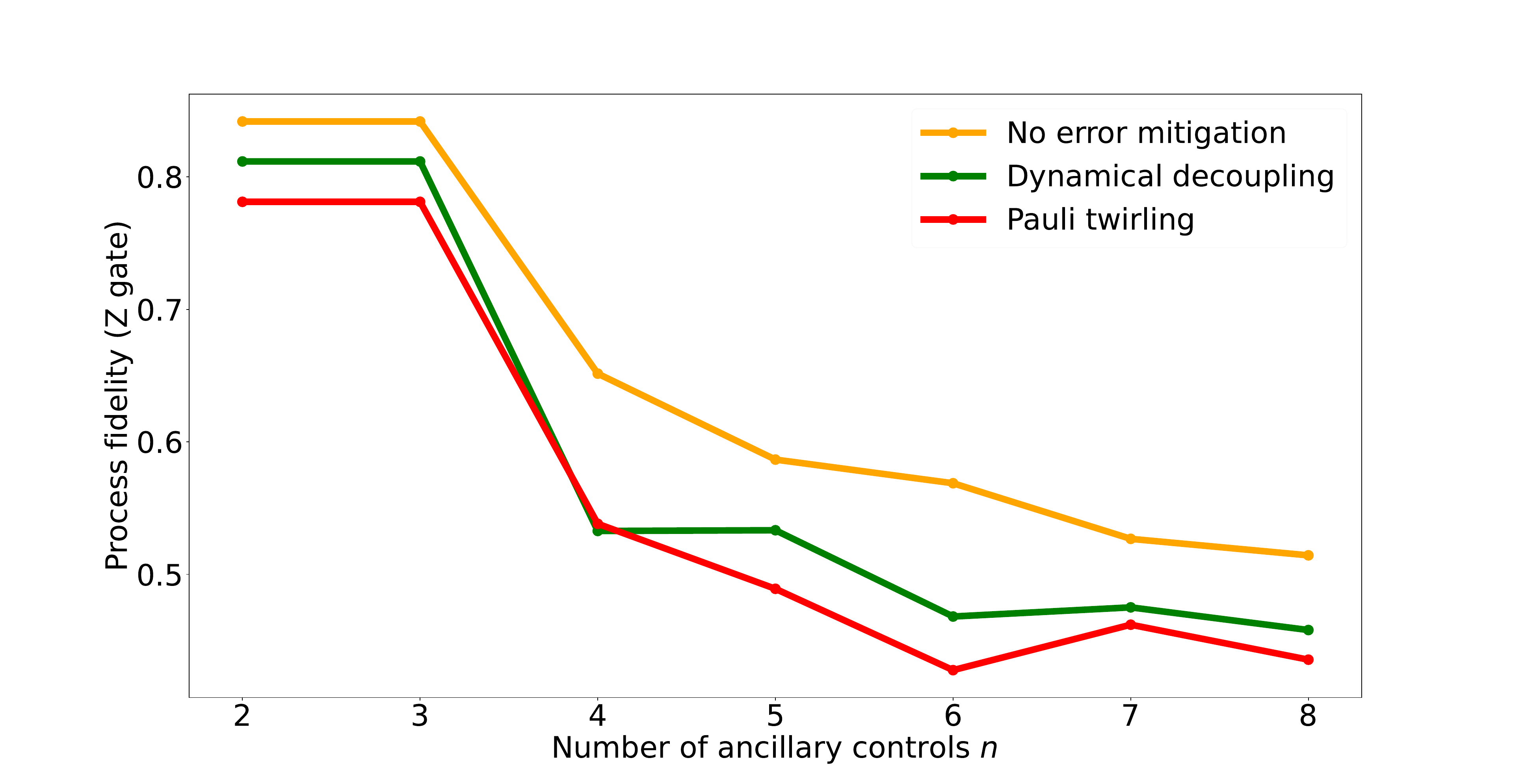}}
	\caption{Data from live runs for $T^*$ with $n=2$ to $n=8$. The circuit for $n=3$ reduces to that for $n=2$.}
	\label{fig:PlotTorino}
\end{figure}
\begin{figure}[tbp]
	\subcaptionbox{Probability of applying $\sqrt{T}^*$, live runs.\label{fig:PlotTorinoProbTPi8}\medskip}{\includegraphics[width=1\linewidth,trim=125 25 125 114,clip]{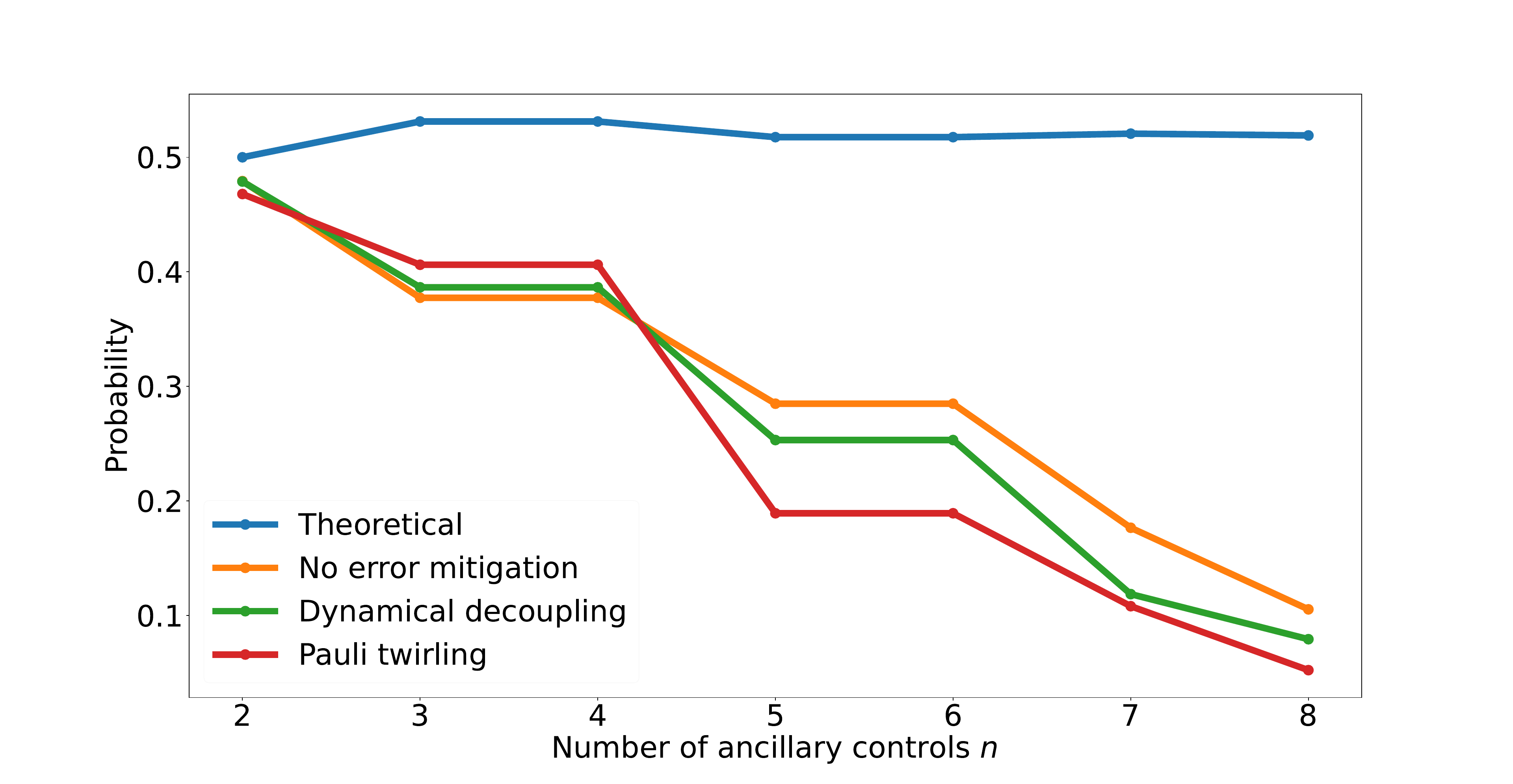}}
	\subcaptionbox{PF for $\sqrt{T}^*$, live runs.\label{fig:PlotTorinoPFTPi8}\medskip}{\includegraphics[width=1\linewidth,trim=105 25 125 114,clip]{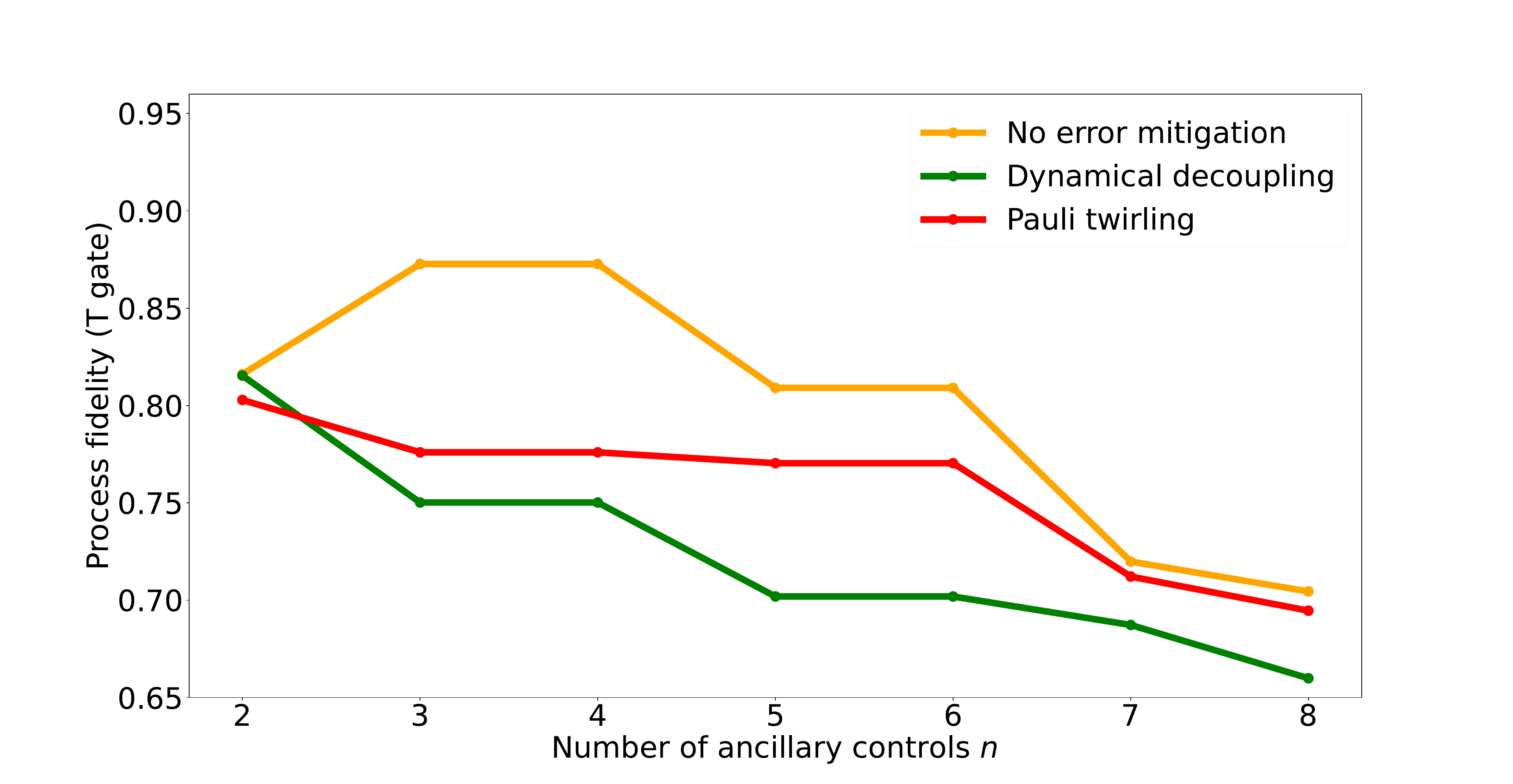}}
	\subcaptionbox{PF for $Z^*$ ($\theta = \pi/8$), live runs.\label{fig:PlotTorinoPFZPi8}}{\includegraphics[width=1\linewidth,trim=125 25 125 114,clip]{PlotTorinoPFZ}}	
  \caption{Data from live runs for $\sqrt T^*$ with $n=2$ to $n=8$. The circuit for $n=4$ reduces to that for $n=3$ and the circuit for $n=6$ reduces to that for $n=5$. The circuit for $n=2$ reduces to the circuit given by $n=1$, as seen in Figure \ref{fig:n2_pi8} in Appendix \ref{app:CircuitsPi8}.}
	\label{fig:PlotTorinoPi8}
\end{figure}
The results of the live runs, together with a graph of theoretical probabilities, are shown in Figures \ref{fig:PlotTorino} and \ref{fig:PlotTorinoPi8}.

The graphs show results after each circuit is iterated $10^5$ times (for a specific $n$, this corresponds to $12\cdot 10^5$ iterations).
We begin by looking at live run results for $\theta = \frac{\pi}{4}$. 
We see that the probabilities reasonably match the simulated results in Figure \ref{fig:PlotSimProb}, having a high probability for $n=2$ and $n=3$, and a rapid decline for $n=4$ followed by a slow decrease (akin to that of $\delta=0.1$ in Figure \ref{fig:PlotSimProb}).
The results for $n=2$ and $n=3$ are fairly similar, with the graph for Pauli twirling barely changing, which is reasonable as the algorithm outputs the same $k$ and $n$ values, resulting in the same circuit.
This is further exemplified in Figure \ref{fig:PlotTorinoProbTPi8}, where $n=3$ and $n=4$ give similar results since the same circuit is implemented, and the same applies to $n=5$ and $n=6$.

The live run PFs, as seen in Figures \ref{fig:PlotTorinoPFT} and \ref{fig:PlotTorinoPFTPi8}, match the simulation results well.
The graphs show behavior similar to that of the probabilities when a $T^*$ rotation is applied, having initially high values for $n=2$ and $n=3$, followed by a slight decline after $n=3$, where the PF for $T^*$ rotation remains high as $n$ is increased.
For $\sqrt{T}^*$ rotation, the PF is high for all $n$.
This matches the expected results well, as we expect the PFs to be constant regardless of $n$, as seen in Figure \ref{fig:PlotSimPFT}.
When a $T^*$ rotation is applied, it is difficult to determine which error mitigation method gives the best results, as the results for all methods oscillate between each other, while for $\sqrt{T}^*$ rotation no error mitigation gives the best results.
The PF for $Z^*$ rotation, as seen in Figures \ref{fig:PlotTorinoPFZ} and \ref{fig:PlotTorinoPFZPi8}, also matches the expected results, having an initially high PF which gradually decreases towards 0.5 as $n$ is increased.

\subsection{Modeling hardware noise using simulated and live run results} \label{sec:sim_live_comparison}

\begin{figure}[tbp]
	\subcaptionbox{Rotation probability for different noise models, given $\theta = \frac{\pi}{4}$.\label{fig:ErrorProb}\medskip}{\includegraphics[width=1\linewidth,trim=125 35 125 114,clip]{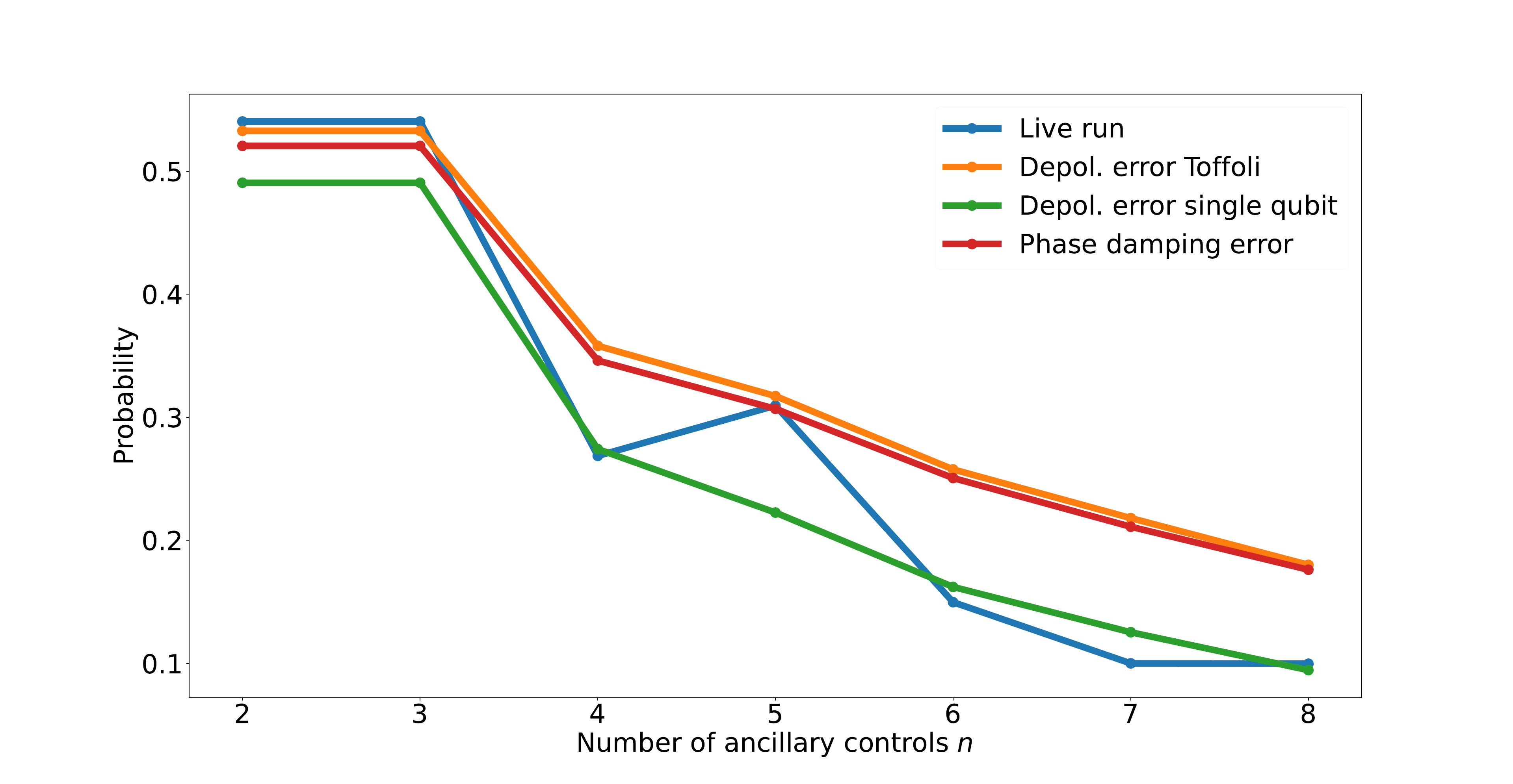}}
	\subcaptionbox{Process fidelity for different noise models, given $\theta = \frac{\pi}{4}$.\label{fig:ErrorProbPFT}\medskip}{\includegraphics[width=1\linewidth,trim=105 35 125 114,clip]{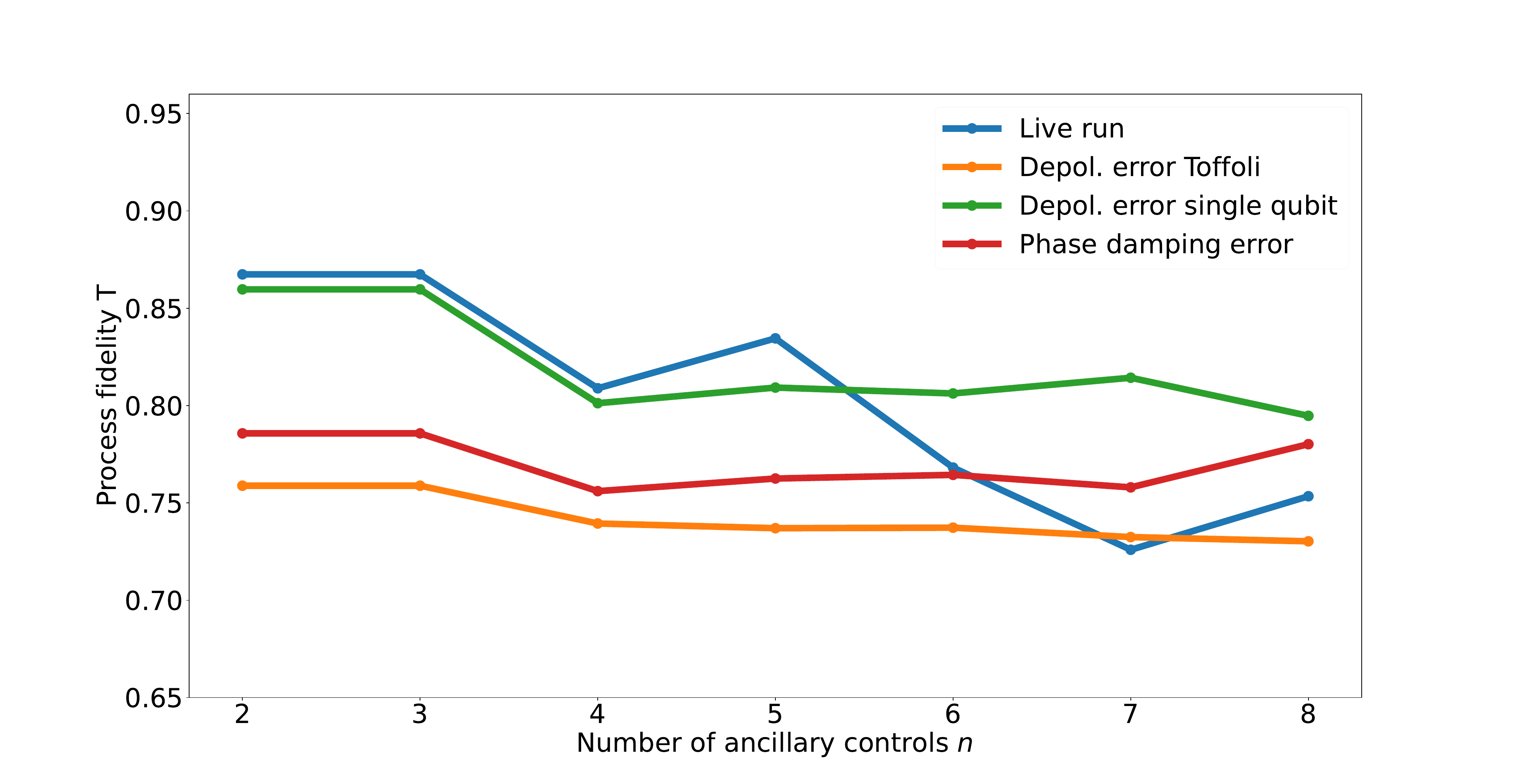}}
	\caption{Rotation probability and process fidelity for different error models. }
	\label{fig:ErrorProbs}
\end{figure}
\begin{figure}[tbp]
	\subcaptionbox{Comparison of rotation probabilities. \label{fig:OptimalComparisonProb}\medskip}{\includegraphics[width=1\linewidth,trim=125 25 125 114,clip]{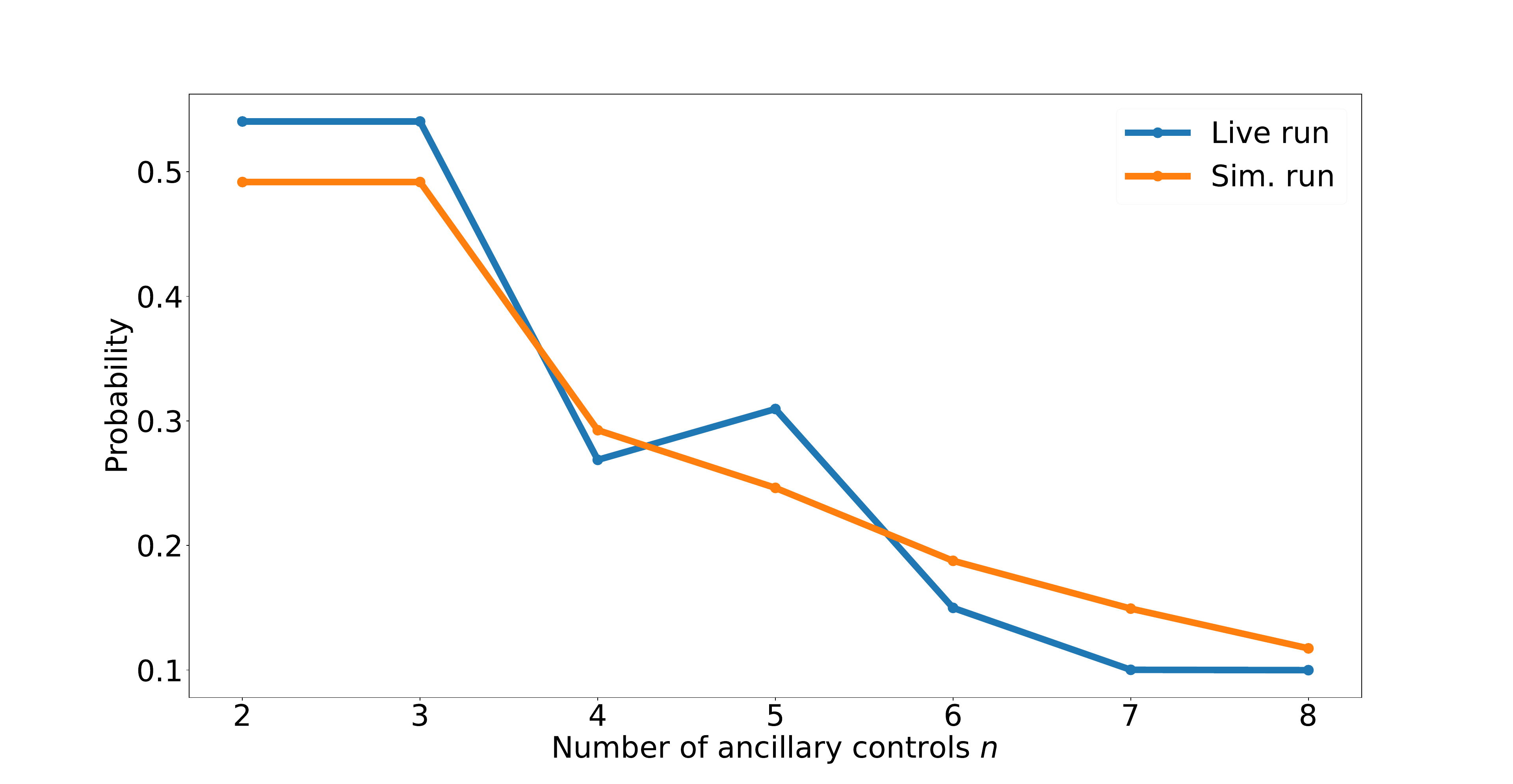}}
	\subcaptionbox{Comparison of process fidelities. \label{fig:OptimalComparisonPFT}\medskip}{\includegraphics[width=1\linewidth,trim=100 25 125 114,clip]{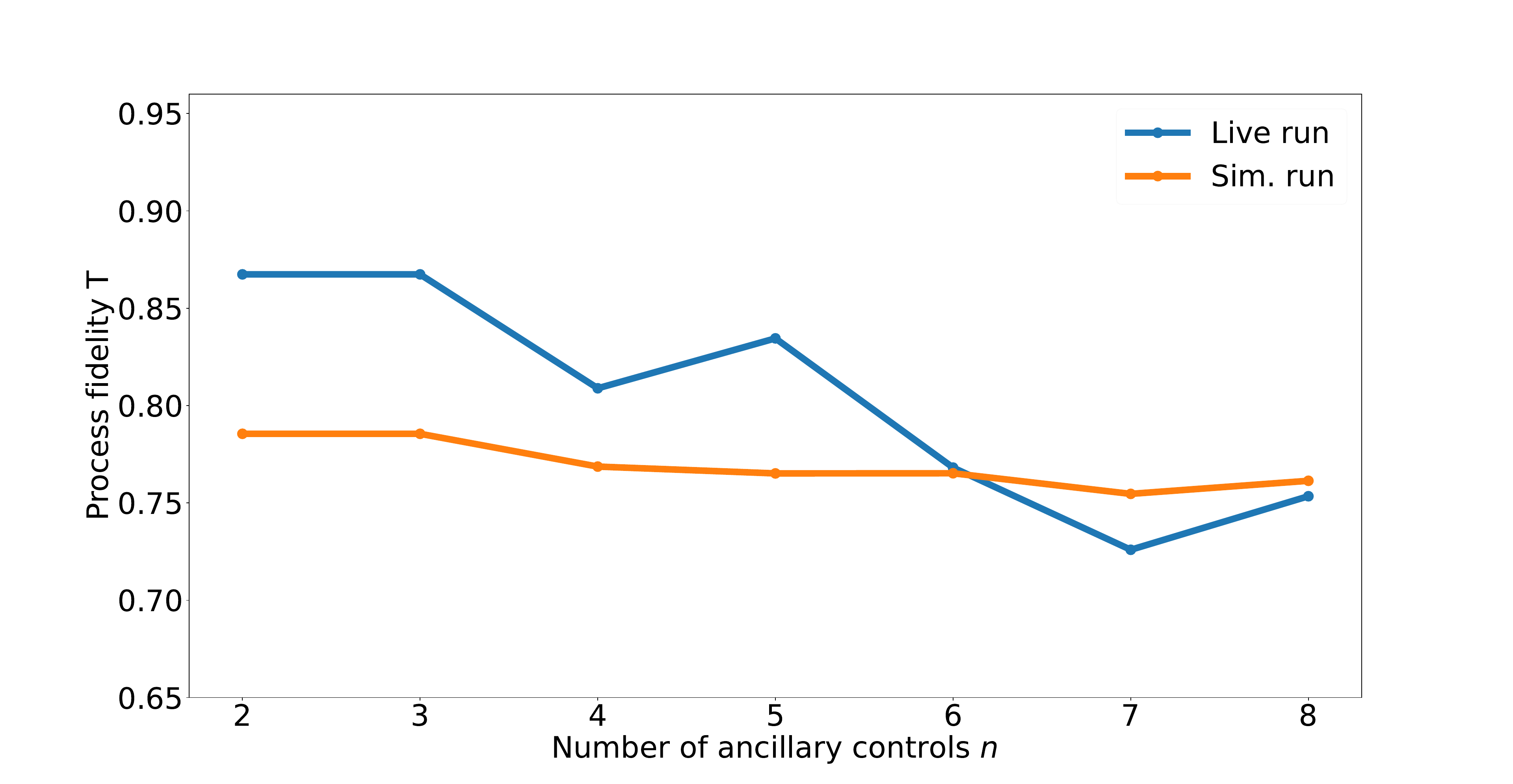}}
	\caption{Comparison of rotation probabilities and process fidelities for live runs and simulations using optimal error parameters, given $\theta = \frac{\pi}{4}$.}
	\label{fig:OptimalComparisons}
\end{figure}
\begin{figure}[tbp]
	\subcaptionbox{Comparison of rotation probabilities. \label{fig:OptimalComparisonProbPi8}\medskip}{\includegraphics[width=1\linewidth,trim=100 25 125 114,clip]{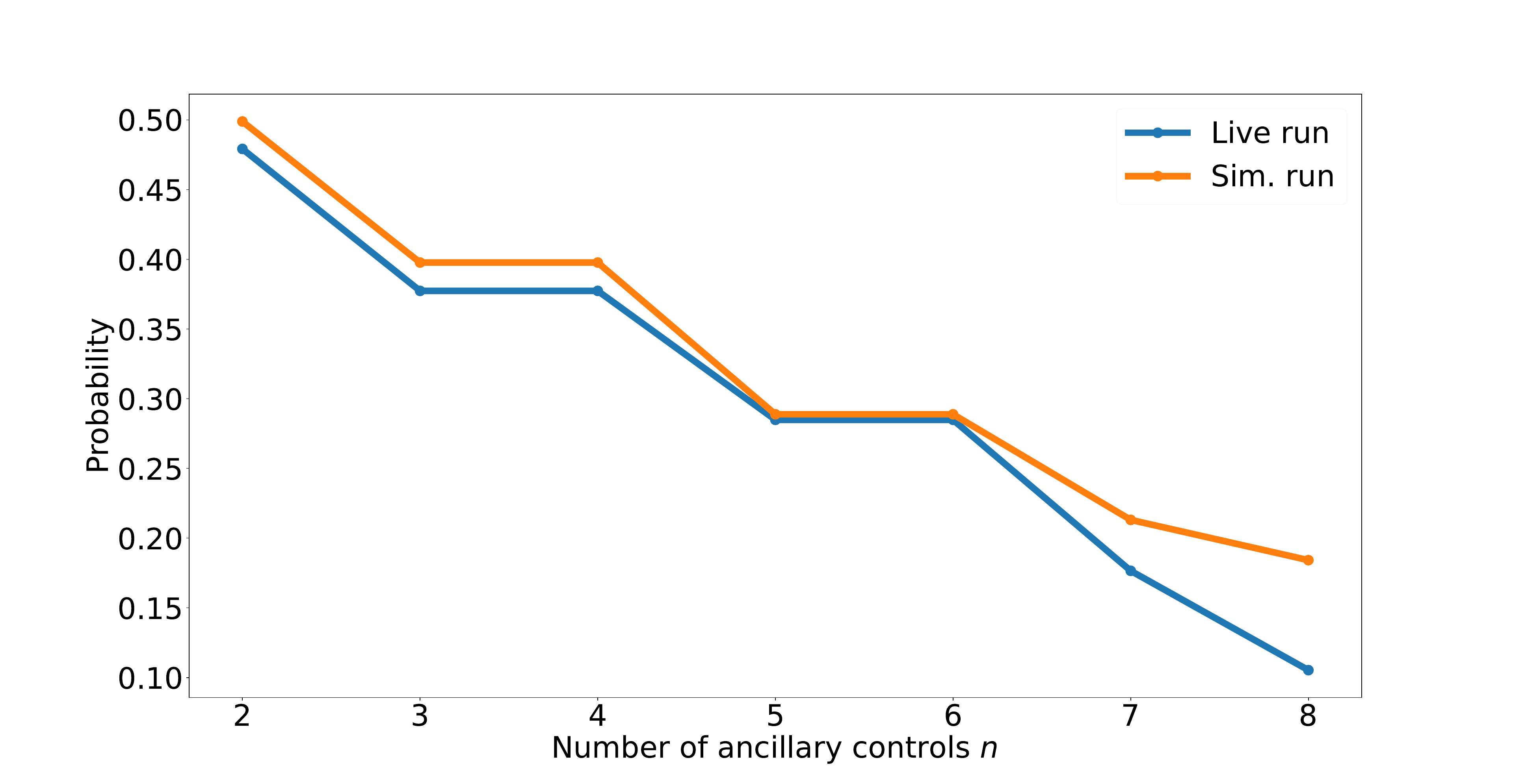}}
	\subcaptionbox{Comparison of process fidelities. \label{fig:OptimalComparisonPFTPi8}\medskip}{\includegraphics[width=1\linewidth,trim=100 25 125 114,clip]{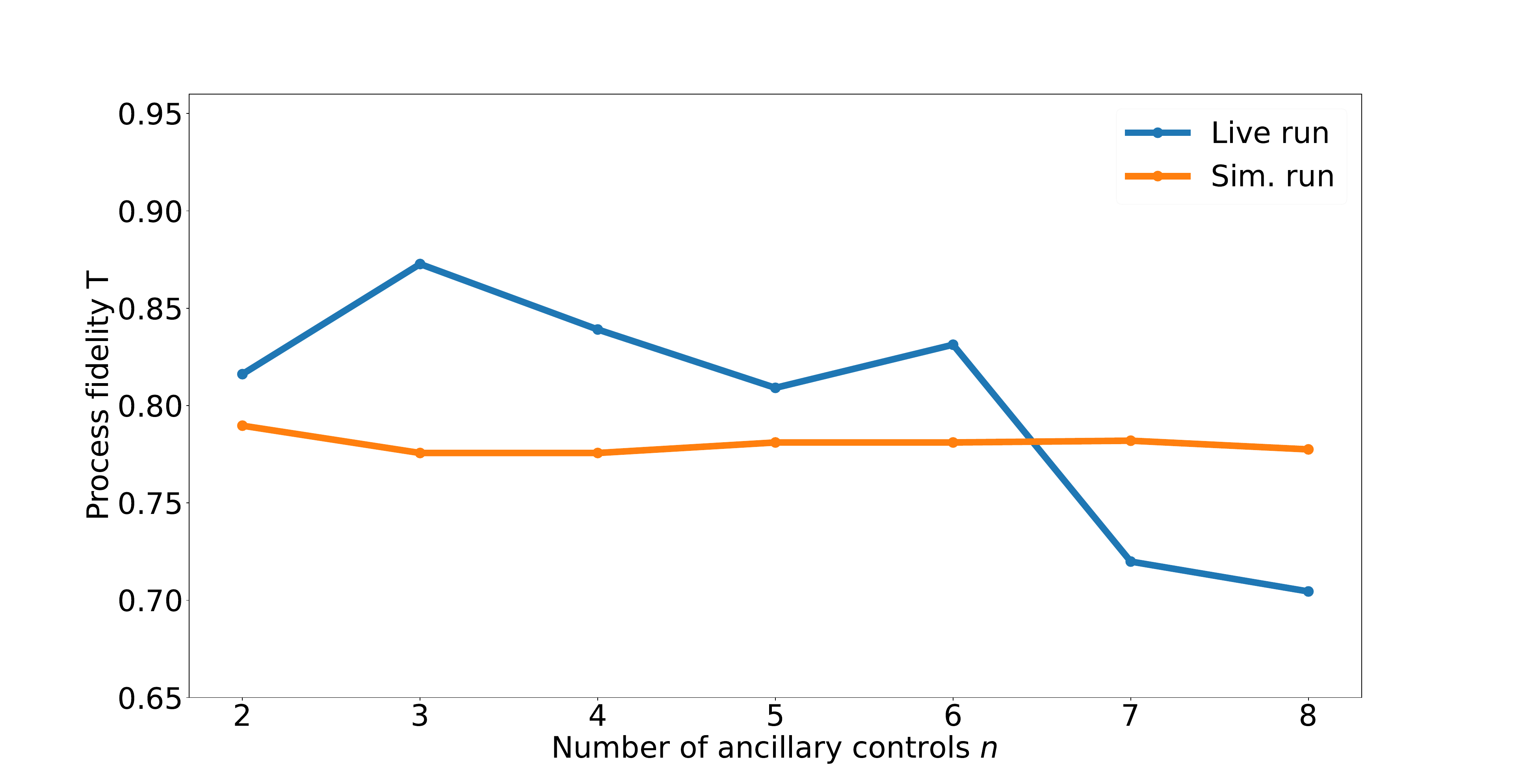}}
	\caption{Comparison of rotation probabilities and process fidelities for live runs and simulations using optimal error parameters, given $\theta = \frac{\pi}{8}$.}
	\label{fig:OptimalComparisonsPi8}
\end{figure}

Here we attempt to characterize the noise in the hardware using the live run results.
QISKIT offers simulation of a handful of quantum error models, of which we have elected to use depolarization for single-qubit and Toffoli gates separately, and phase damping. 
Amplitude damping is less suited for the present system, since it is best suited to model photon loss in transmission lines, which is not the physical system we use here, and the remaining models, such as thermal relaxation, add a large set of parameters.
The latter would make optimization very complex, and some of these parameters can be estimated only with knowledge of the physical setup, which we do not have.
Figures \ref{fig:ErrorProb} and \ref{fig:ErrorProbPFT} show probabilities and PFs for $T^*$ rotation when different noise models are applied separately with different error parameters $\delta_i$ for each respective noise model, while for the live run data no error mitigation is used.
The $\delta_i$ values were adjusted so that the simulated results fit the live run data as well as possible for all $n$ with the use of a least-squares fit, in order to show how each respective noise model behaves. 
The error parameters seen in Figure \ref{fig:ErrorProbs} are as follows: 0.2 depolarizing error applied to Toffoli gates, 0.08 depolarizing error applied to single qubits, and 0.35 phase error.
In Figure \ref{fig:ErrorProb}-\ref{fig:ErrorProbPFT} we see that none of the models match the live run results jointly for probability and PF. 
The error models that give higher rotation probabilities than the live runs also give lower PF compared with the live runs.
Changing the error rates is not useful as this simply shifts both graphs along the y-axis in the same direction.

It is therefore reasonable to assume that by combining the noise models and varying $\delta_i$, we will be able to find configurations where both probabilities and PF match the live run results. 
Doing this is also reasonable as we expect there to be different sources of noise in the quantum hardware. 
We therefore attempted to combine these error models, while varying their respective error rates.
We also decided to have two separate error rates when applying depolarizing error to single qubits and to Toffoli gates as it is reasonable to assume that the Toffoli operation would introduce more noise in the hardware, given the implementation complexity.
Simulations were performed for $\theta = \frac{\pi}{4}$ and $\theta = \frac{\pi}{8}$, where we picked optimal $\delta_i$ that fit the simulation data for both angles.

To find the best matching error rates, we used the method of least squares to minimize the distance between the live run data and simulated data, with equal weight of the probability and the PF of $T^*$ rotation.
The optimal results are shown in Figures \ref{fig:OptimalComparisonProb} and \ref{fig:OptimalComparisonPFTPi8}, where we get the following optimal parameters for $T^*$ rotation: 0.13 depolarizing error applied to Toffoli gates, 0.03 depolarizing error for single qubits, and 0.05 phase damping error. 
In a physical device, these parameters would correspond to a $90\%$ PF for a NOT gate.
For $\sqrt{T}^*$ rotation we get 0.06 depolarizing error applied to Toffoli gates, 0.05 depolarizing error for single qubits, and 0.02 phase damping error.
From Figures \ref{fig:OptimalComparisonProb} and \ref{fig:OptimalComparisonProbPi8} we can see a much better match for both $\theta = \frac{\pi}{4}$ and $\theta = \frac{\pi}{8}$.
We get slight variations from the live runs but the overall shape of the simulated graphs matches well.

\section{Conclusions}
We have devised an experimental implementation of a one-shot version of the single-qubit rotation algorithm described in Ref.~\cite{Hindlycke2024a}, and carried out experimental runs on both simulated (at various noise levels) and real quantum computers using the QISKIT software suite \cite{JavadiAbhari2024}.
We provide a full description of the simplifications of the circuit used in the single-qubit rotation algorithm, and introduce several additional ones, as reducing circuit depth and gate count is always useful in practical implementations.
Simulation outcomes are well in line with theoretical predictions and suggest some simple patterns, which provides motivation for further study.
Live run outcomes adhere less well to predictions, although they have a form similar to that of the simulated results.

We used several error models in our simulations to model the noise experienced using the hardware.
The combination of three error models allows us to approximate the noise fairly well, indicating that our algorithm works under modest noise levels and that present hardware has decent performance.

The live runs show promising results for low values of $n$, and testing higher values of $n$ is simple as the Toffoli count and gate depth of our algorithm increases linearly with $n$ and only slight adjustments to the circuit are required.
When quantum computer performance is being assessed, this gives an advantage over algorithms such as Shor's algorithm which have a far higher gate depth even in simple examples, at least those that are nontrivial \cite{Smolin2013}. 
We also note that Shor's algorithm is qubic in complexity rather than linear.
This suggests that our algorithm would be useful as a benchmark for the rapidly developing field of quantum computation, and could be used to assess quantum computing hardware both in NISQ devices and in the early FTQC era.

\section{Acknowledgments}
We acknowledge support from the WACQT Quantum Technology Testbed operated by Chalmers Next Labs, which is funded by the Knut and Alice Wallenberg Foundation.
This project was supported by the Swedish Science Council (Project No. 2023-05031).

\section{Data availability}
All experimental data is available on reasonable request made to the corresponding author.

\bibliography{lib}
\clearpage

\appendix
\section{Qiskit circuits given $\theta=\frac{\pi}{4}$}
\label{app:Circuits}
\begin{figure}[H]
  \centering
	\includegraphics[scale=0.35]{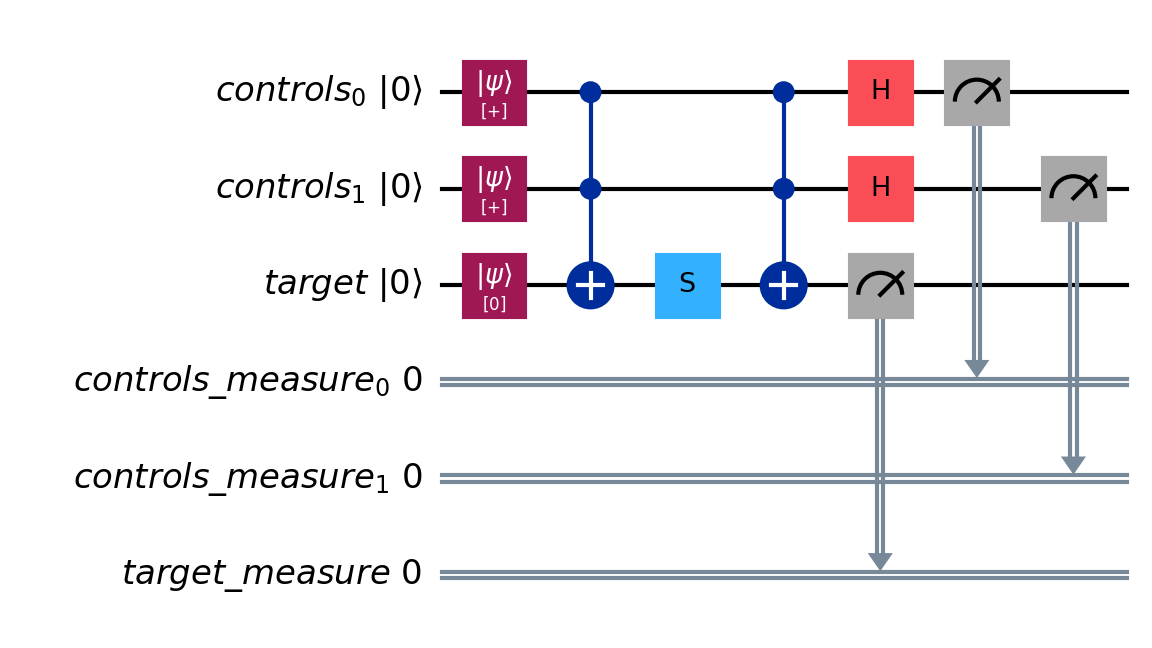}
	\caption{Circuit implementing $R_{\arccos 3/5}$, $n = 2$ and $n =3$.}
	\label{circ:pi4nc}
\end{figure}

\begin{figure}[H]
  \centering
	\includegraphics[width=.9\linewidth]{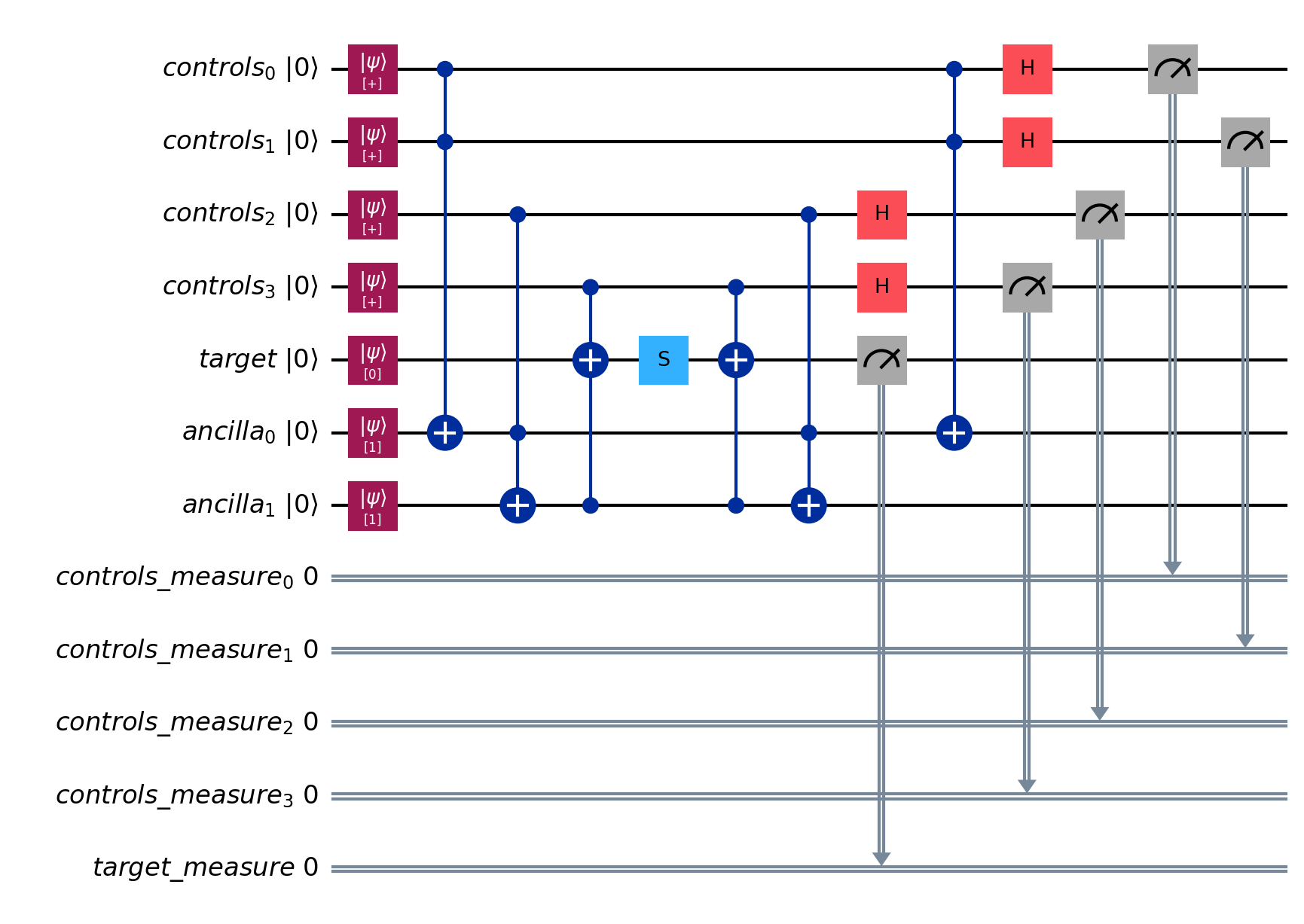}
	\caption{Single-shot rotation algorithm circuit, $n = 4$.}
	\label{circ:pi4n4}
\end{figure}

\begin{figure}[H]
  \centering
	\includegraphics[width=.9\linewidth]{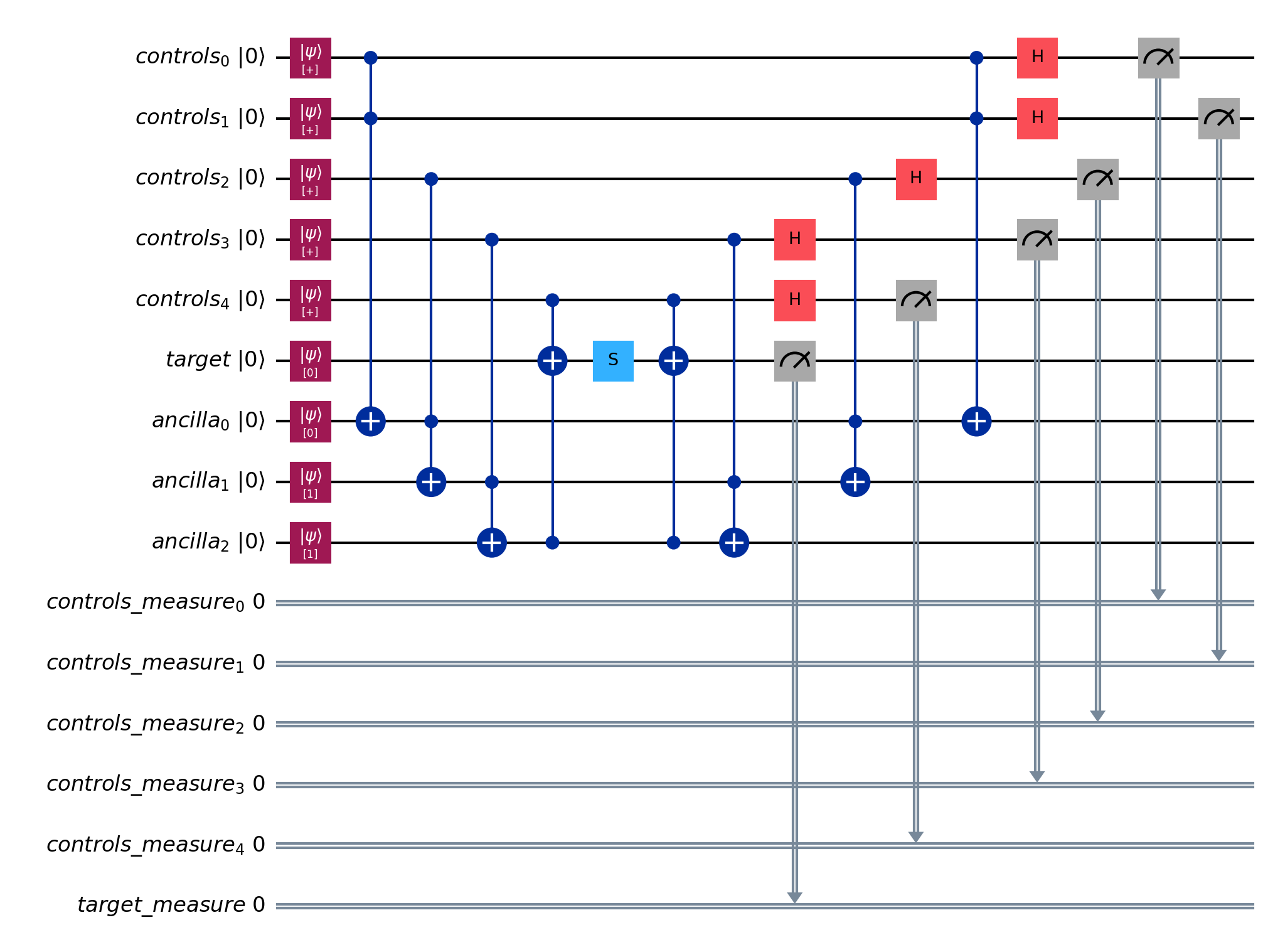}
	\caption{Single-shot rotation algorithm circuit, $n = 5$.}
	\label{circ:pi4n5}
\end{figure}

\begin{figure}[H]
  \centering
	\includegraphics[width=.9\linewidth]{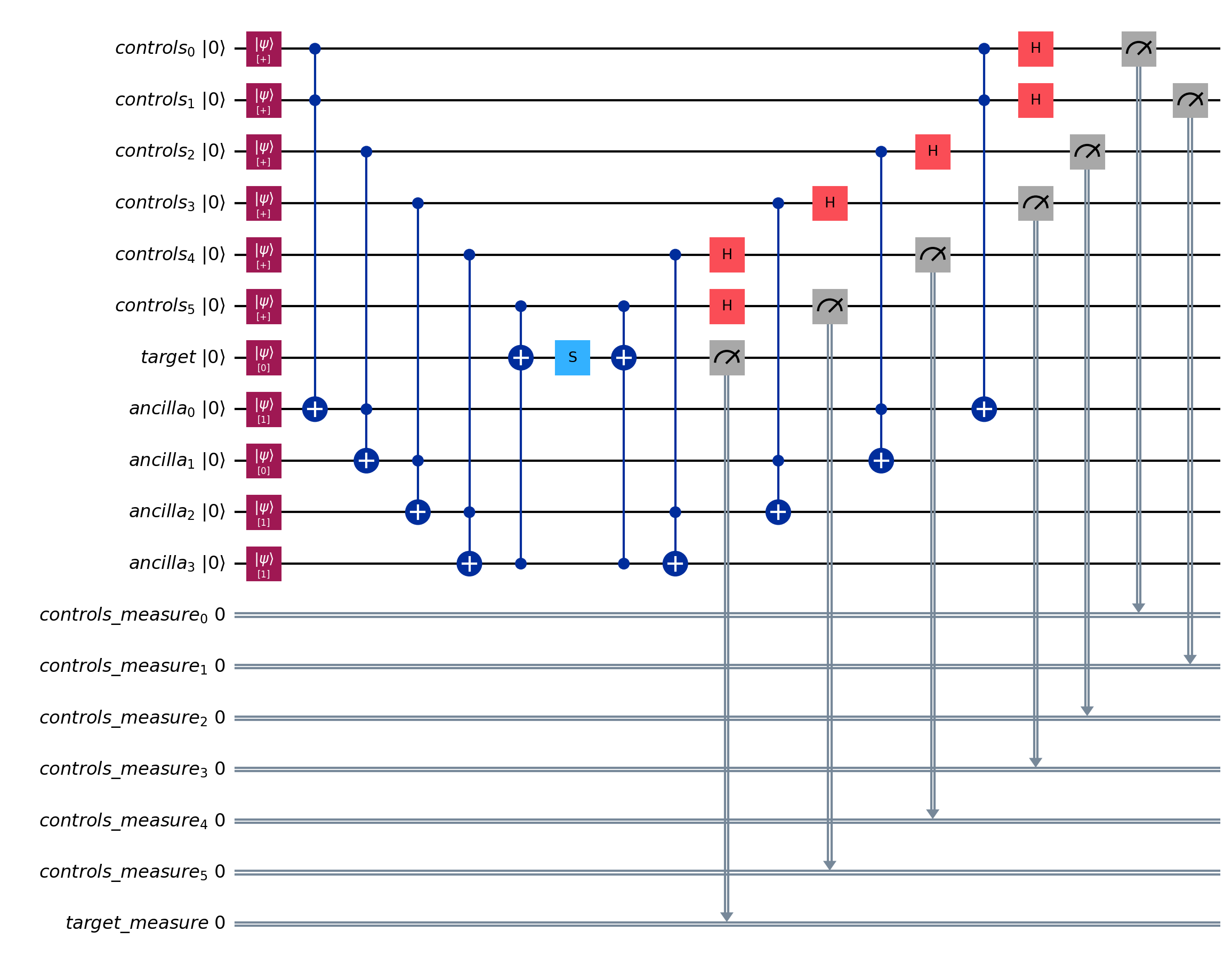}
	\caption{Single-shot rotation algorithm circuit, $n = 6$.}
	\label{circ:pi4n6}
\end{figure}

\begin{figure}[H]
  \centering
	\includegraphics[width=.9\linewidth]{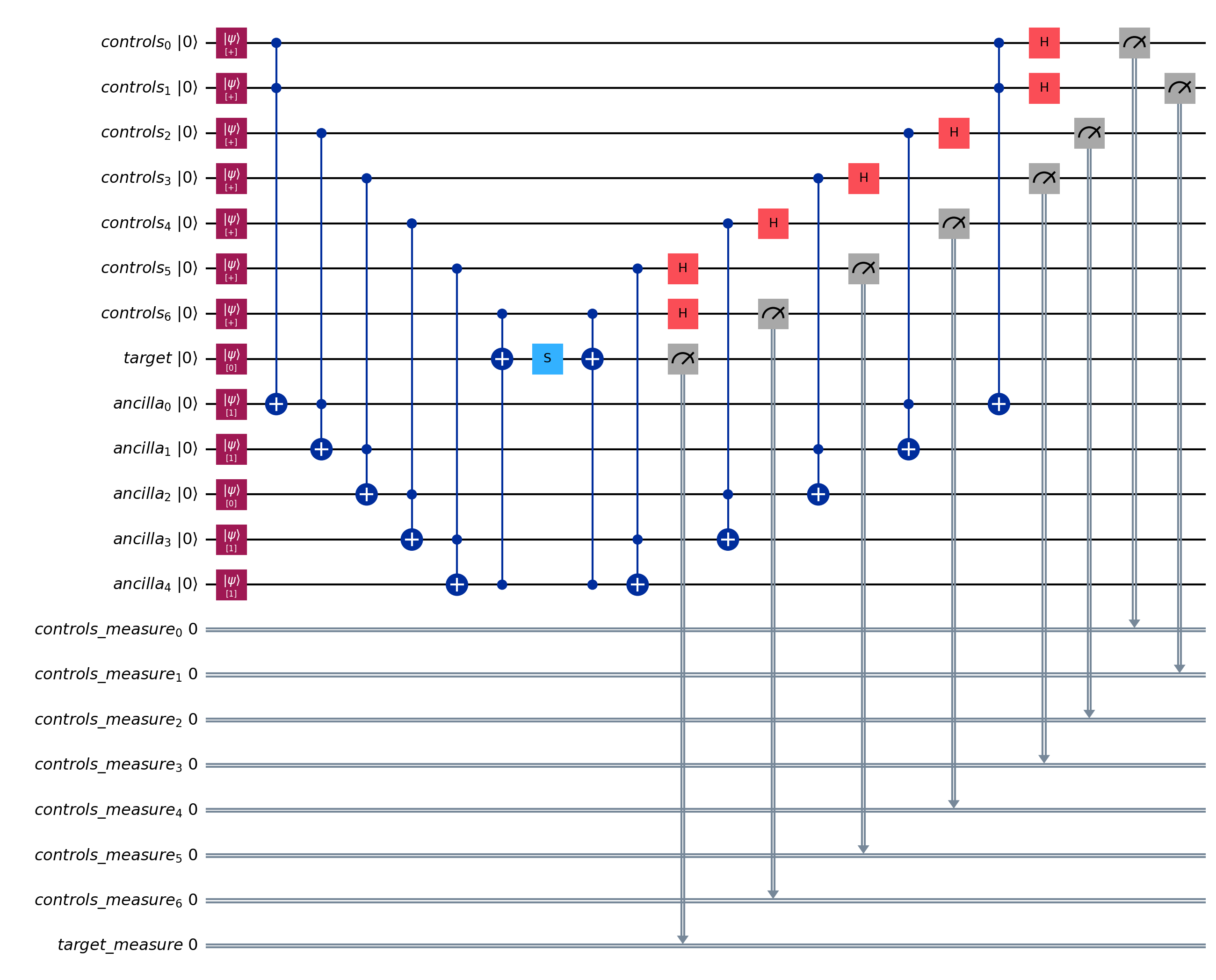}
	\caption{Single-shot rotation algorithm circuit, $n = 7$.}
	\label{circ:pi4n7}
\end{figure}

\begin{figure}[H]
  \centering
	\includegraphics[width=.9\linewidth]{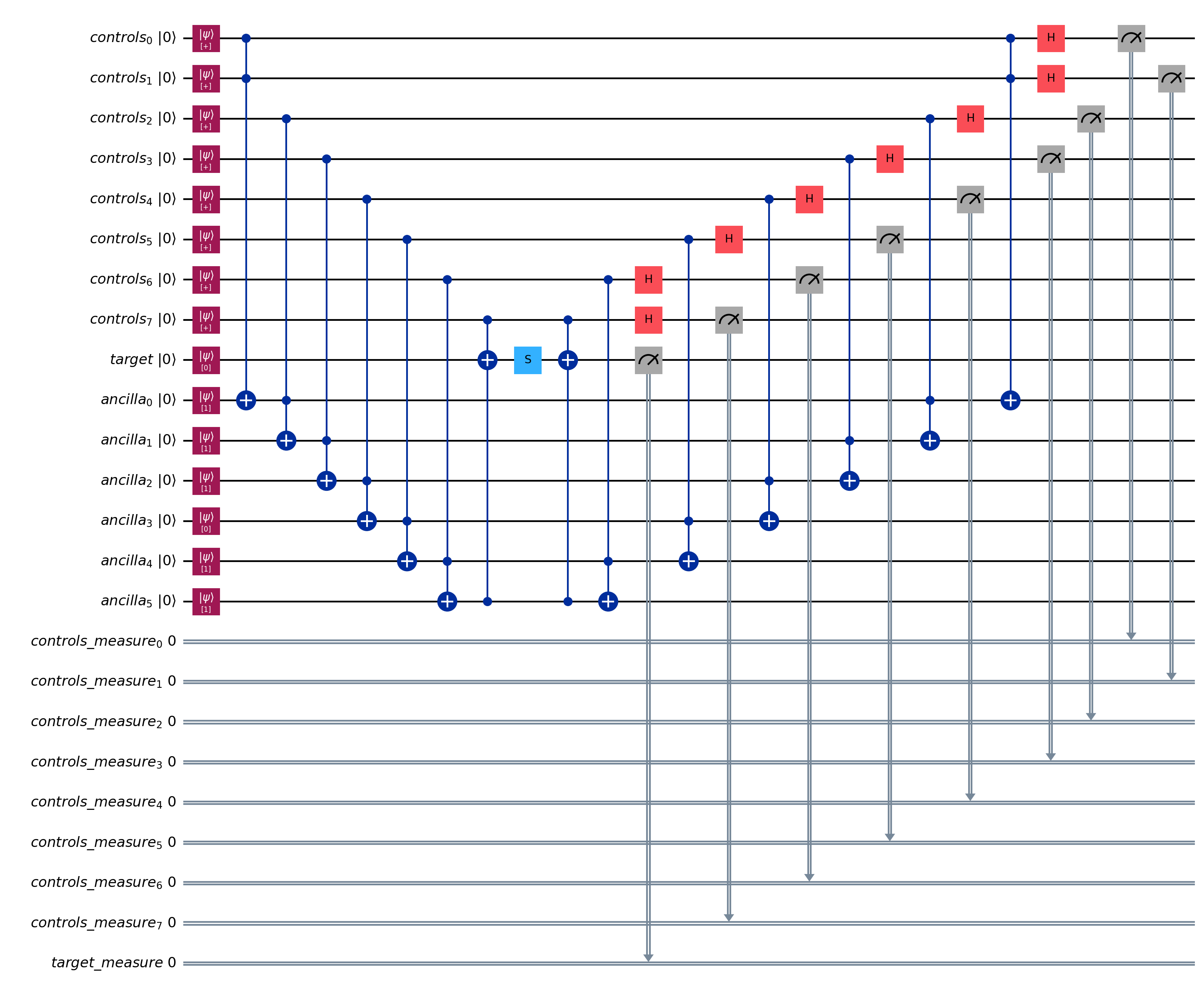}
	\caption{Single-shot rotation algorithm circuit, $n = 8$.}
	\label{circ:pi4n8}
\end{figure}

Figures \ref{circ:pi4nc} - \ref{circ:pi4n8} show (pretranspiling) circuits used for $\theta = \tfrac{\pi}{4}$ in this work.
Note that here the target qubit is in the state $\ket{0}$ and measurements are done in the Pauli $Z$ basis; in addition during the experiment, input states $\ket{1}, \ket{+},$ and $\ket{+i}$ and Pauli $X$ and Pauli $Y$ measurement bases were used.

\clearpage

\section{Qiskit circuits given $\theta=\frac{\pi}{8}$}
\label{app:CircuitsPi8}

\begin{figure}[H] 
	\centering
	\includegraphics[scale=0.35]{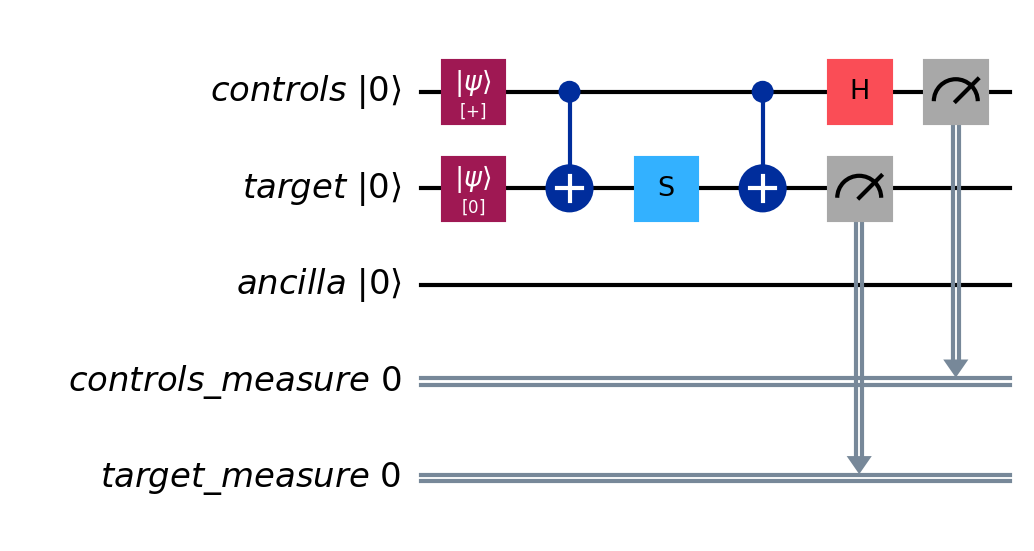}
	\caption{Single-shot rotation algorithm circuit, $n = 2$.}
	\label{fig:n2_pi8}
\end{figure}

\begin{figure}[H]
	\centering
	\includegraphics[width=.9\linewidth]{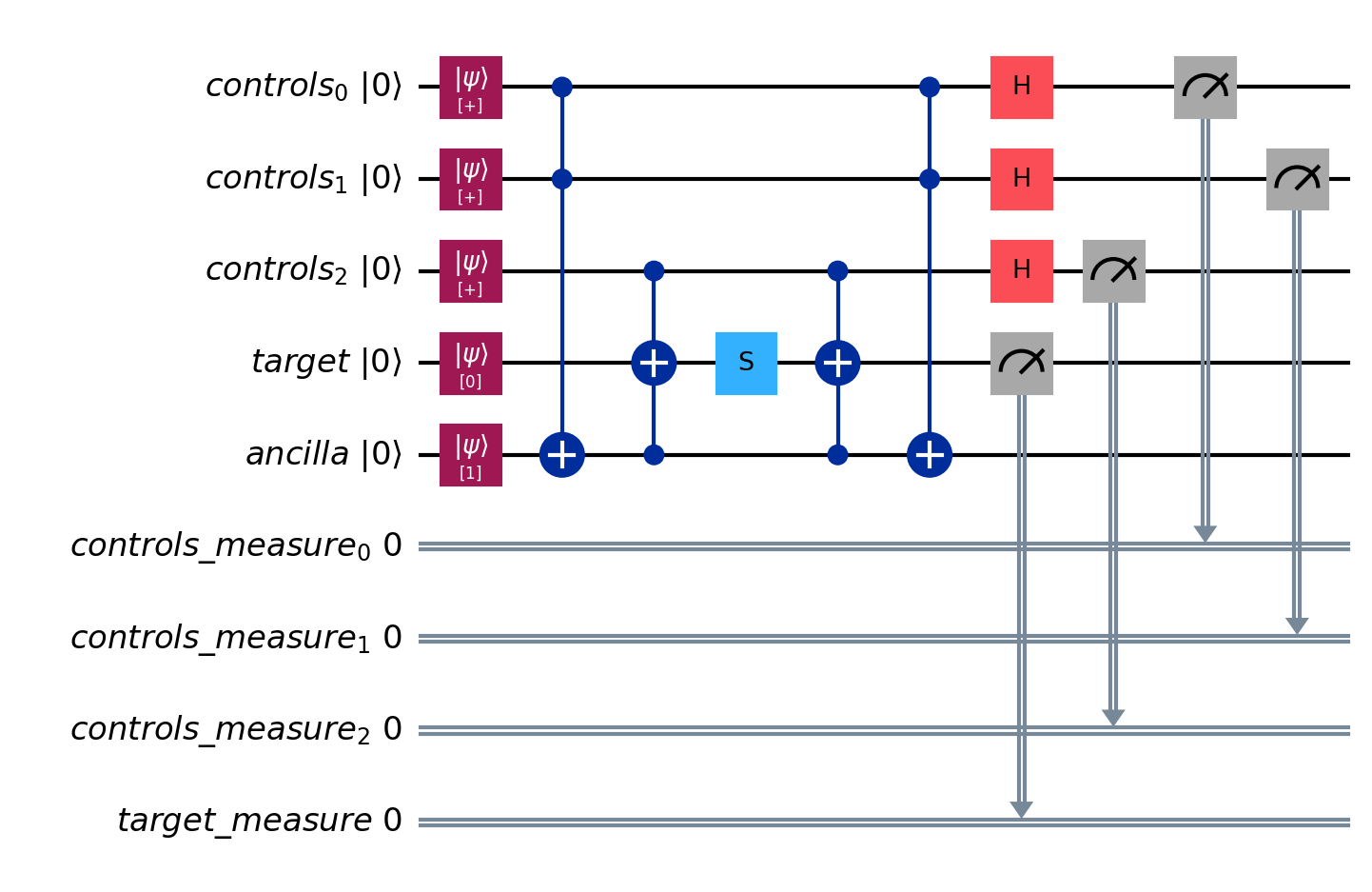}
	\caption{Single-shot rotation algorithm circuit, $n = 3$ and $n=4$.}
	\label{circ:pi8n34}
\end{figure}

\begin{figure}[H]
	\centering
	\includegraphics[width=.9\linewidth]{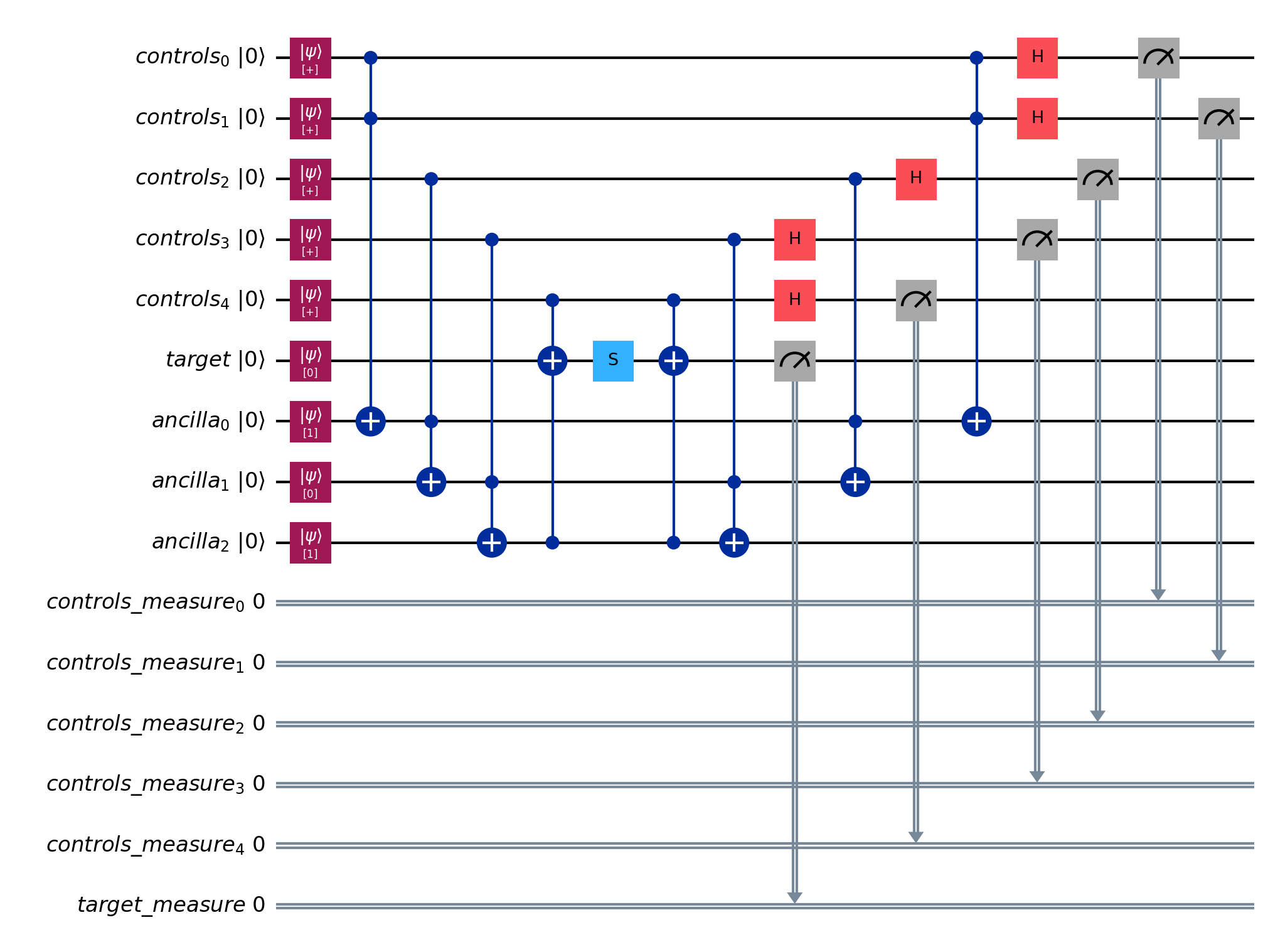}
	\caption{Single-shot rotation algorithm circuit, $n = 5$ and $n=6$.}
	\label{circ:pi8n56}
\end{figure}

\begin{figure}[H]
	\centering
	\includegraphics[width=.9\linewidth]{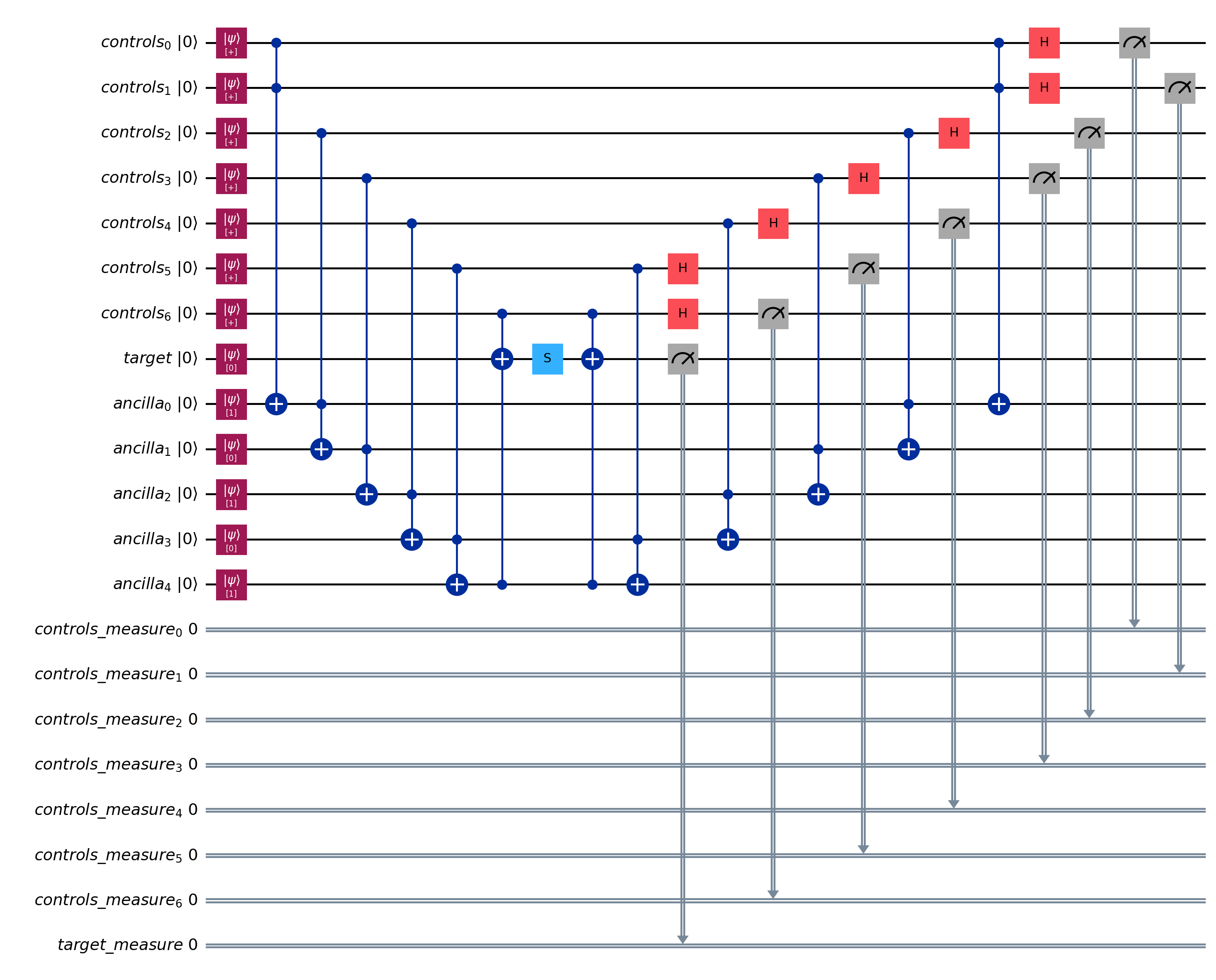}
	\caption{Single-shot rotation algorithm circuit, $n = 7$.}
	\label{circ:pi8n7}
\end{figure}

\begin{figure}[H]
	\centering
	\includegraphics[width=.9\linewidth]{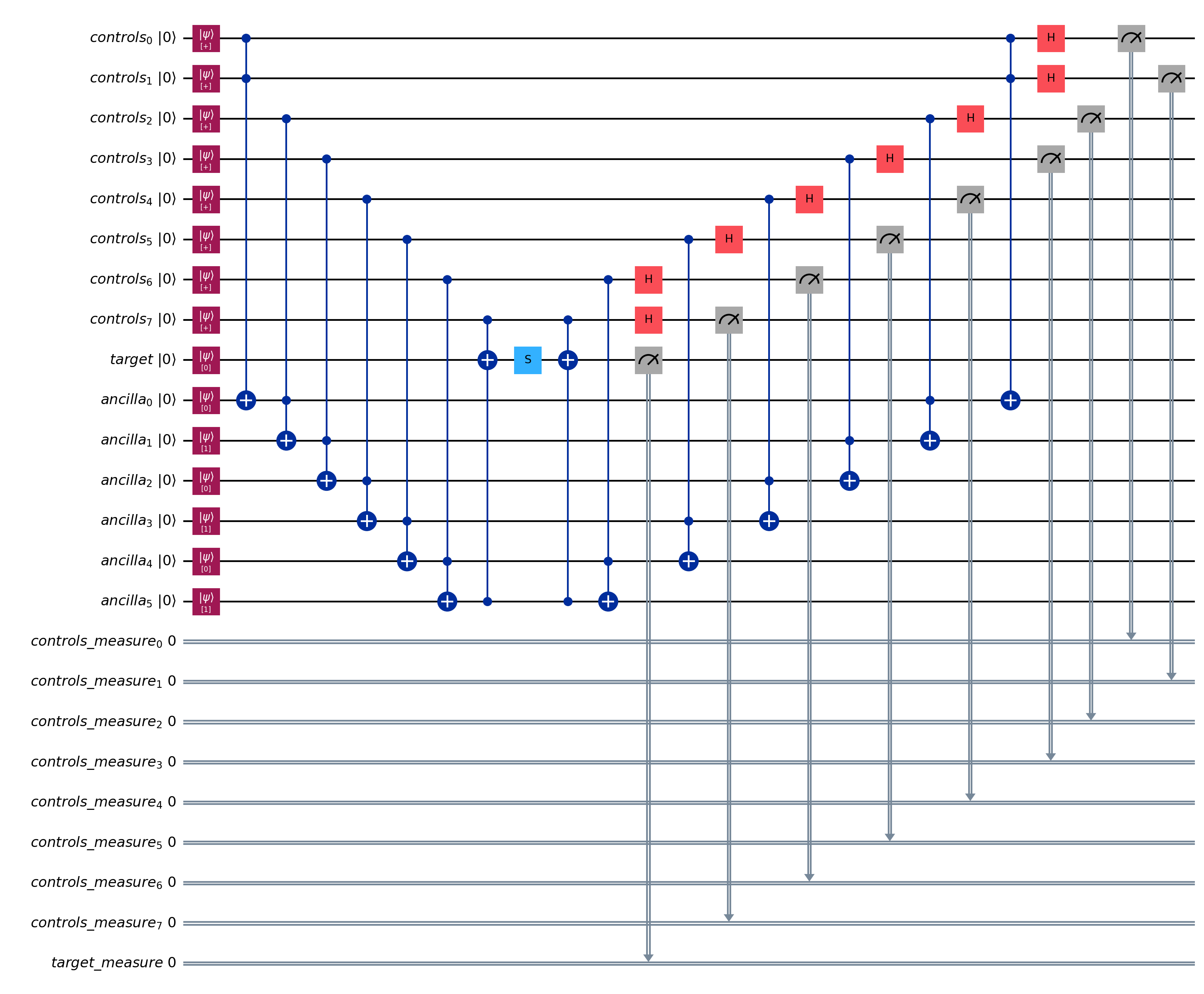}
	\caption{Single-shot rotation algorithm circuit, $n = 8$.}
	\label{circ:pi8n8}
\end{figure}

Figures \ref{fig:n2_pi8} - \ref{circ:pi8n8} show (pretranspiling) circuits used for $\theta = \tfrac{\pi}{8}$ in this work.
Here again the target qubit is in the state $\ket{0}$ and measurements are done in the Pauli $Z$ basis;
also, input states $\ket{1}, \ket{+},$ and $\ket{+i}$ and Pauli $X$ and Pauli $Y$ measurement bases were used.

\clearpage

\section{Tables}
\label{app:Tables}

Tables \ref{tbl:prob} - \ref{tbl:live_paulitwirling} contain the simulated and live run results used for generating Figures \ref{fig:ExactAnglesProb} - \ref{fig:PlotTorinoPi8}.

\footnotesize
\begin{table}[H]
	\centering
	\setlength{\tabcolsep}{1ex}
	\begin{tabular}{ccc}
		\toprule n&Prob $T^*$&Prob $\sqrt{T}^*$\\ \midrule
		2&0.62500&0.50000\\
		3&0.62500&0.53125\\
		4&0.57031&0.53125\\
		5&0.59570&0.51757\\
		6&0.58252&0.51757\\
		7&0.58899&0.52062\\
		8&0.58572&0.51907\\
		\bottomrule
	\end{tabular}
	\caption{Theoretical probability of applying $T^*$ and $\sqrt{T}^*$ rotations.}
	\label{tbl:prob}
\end{table}

\begin{table}[H]
\centering
	\begin{tabular}{ccccc}
		\toprule $\delta$&$n$&Prob&PF $T^*$&PF $Z^*$\\ 
    \midrule
		0.0&2&0.62485&0.99386&1.00057\\
		&4&0.57030&1.00052&0.99837\\
		&5&0.59560&1.00086&0.99948\\
		&6&0.58278&0.99996&1.00314\\
		&7&0.58930&0.99989&0.99868\\
		&8&0.58647&0.99891&1.00156\\
		\midrule
    0.01&2&0.60219&0.95678&0.90965\\
		&4&0.52136&0.95413&0.86358\\
		&5&0.52946&0.95783&0.84042\\
		&6&0.50276&0.95716&0.82172\\
		&7&0.49330&0.96146&0.80433\\
		&8&0.47644&0.95717&0.78672\\
		\midrule
    0.05&2&0.52658&0.81281&0.67424\\
		&4&0.37058&0.80210&0.59538\\
		&5&0.33666&0.81079&0.56339\\
		&6&0.28466&0.80832&0.54588\\
		&7&0.25003&0.80814&0.52805\\
		&8&0.21632&0.81024&0.51569\\
		\midrule
    0.1&2&0.45124&0.66806&0.51649\\
		&4&0.25135&0.65267&0.46379\\
		&5&0.20099&0.65507&0.44465\\
		&6&0.14947&0.65720&0.43702\\
		&7&0.11551&0.65195&0.43261\\
		&8&0.08758&0.66398&0.42784\\
		\bottomrule
	\end{tabular}
	\caption{Simulated results.}
	\label{tbl:sim}
\end{table}

\vfill\null

\begin{table}[H]
	\centering
	\begin{tabular}{ccccc}
		\toprule 
    $U$&$n$&Prob&PF $U^*$&PF $Z^*$\\ \midrule
		$T$&2&0.54037&0.86738&0.84180\\
		&3&0.60282&0.93899&0.83488\\
		&4&0.26873&0.80886&0.65150\\
		&5&0.30956&0.83450&0.58662\\
		&6&0.14991&0.76812&0.56878\\
		&7&0.10019&0.72591&0.52680\\
		&8&0.09992&0.75342&0.51436\\
		\midrule
		$\sqrt T$&2&0.47917&0.80198&0.78441\\
		&3&0.37735&0.87270&0.79321\\
		&4&0.43341&0.83906&0.72537\\
		&5&0.28485&0.80909&0.65579\\
		&6&0.30708&0.83124&0.64565\\
		&7&0.17654&0.71987&0.56813\\
		&8&0.10530&0.70451&0.53633\\
		\bottomrule
	\end{tabular}
	\caption{Live run results obtained with no error mitigation.
	}
	\label{tbl:live_noerror}
\end{table}

\begin{table}[H]
	\centering
	\begin{tabular}{ccccc}
		\toprule
    $U$&$n$&Prob&PF $U^*$&PF $Z^*$\\ \midrule
		$T$&2&0.55849&0.90565&0.81163\\
		&3&0.58139&0.85051&0.74665\\
		&4&0.32143&0.74597&0.53276\\
		&5&0.29478&0.79591&0.53330\\
		&6&0.22010&0.75955&0.46818\\
		&7&0.15114&0.76870&0.47510\\
		&8&0.10888&0.77687&0.45799\\
    \midrule
		$\sqrt T$&2&0.47882&0.81538&0.77969\\
		&3&0.38648&0.75020&0.68154\\
		&4&0.42164&0.79009&0.59679\\
		&5&0.25312&0.70192&0.46014\\
		&6&0.23918&0.73823&0.46698\\
		&7&0.11860&0.68730&0.48537\\
		&8&0.07924&0.65992&0.46847\\
		\bottomrule
	\end{tabular}
	\caption{Live run results obtained with dynamical decoupling.
	}
	\label{tbl:live_dynamicaldecoupling}
\end{table}

\begin{table}[H]
	\centering
	\begin{tabular}{ccccc}
		\toprule 
    $U$&$n$&Prob&PF $U^*$&PF $Z^*$\\ 
    \midrule
		$T$&2&0.56536&0.90975&0.78117\\
		&3&0.56273&0.91948&0.79041\\
		&4&0.31947&0.80814&0.53817\\
		&5&0.23515&0.79094&0.48910\\
		&6&0.13587&0.78422&0.42771\\
		&7&0.07696&0.78701&0.46201\\
		&8&0.05591&0.77105&0.43561\\
    \midrule
		$\sqrt T$&2&0.46792&0.80286&0.77448\\
		&3&0.40616&0.77597&0.60436\\
		&4&0.41370&0.86678&0.70094\\
		&5&0.18920&0.77039&0.49494\\
		&6&0.18370&0.76009&0.49749\\
		&7&0.10806&0.71216&0.44115\\
		&8&0.05223&0.69946&0.41632\\
		\bottomrule
	\end{tabular}
	\caption{Live run results obtained with Pauli twirling.
	}
	\label{tbl:live_paulitwirling}
\end{table}

\clearpage

\end{document}